\documentclass[a4paper,11pt]{article}
\usepackage{jheppub} 
\usepackage{lineno}
\usepackage{graphicx} 
\usepackage{amsmath}
\usepackage{tikz}
\usepackage{caption}
\usepackage{subcaption}
\usepackage{amssymb}
\usepackage{braket}
\usepackage{enumitem}

\title{\boldmath Phase Transition and Censorship Principle for Holographic Casimir Effect}

\author{Zhen-Rui Huang, Rong-Xin Miao}
\affiliation{School of Physics and Astronomy, Sun Yat-sen University,\\
2 Daxue Road, Zhuhai 519082, China}

\emailAdd{huangzhr5@mail2.sysu.edu.cn}
\emailAdd{miaorx@mail.sysu.edu.cn}

\abstract{This paper explores the holographic Casimir effect associated with parallel spherical defects. The gravity dual is dominated by the AdS soliton with connected EOW branes at small widths, transitioning to AdS space with disconnected EOW branes as the width increases. Consequently, the holographic Casimir effect undergoes a first-order phase transition and vanishes in the disconnected phase. This new behavior highlights a significant difference from free theories, holographic parallel plane defects, and hyperbolic defects. Additionally, we confirm that the free theories adhere to the holographic bound for the Casimir effect in the case of parallel spherical defects.

Interestingly, we observe that cosmic censorship provides a holographic interpretation of the attractive nature of the Casimir force when identical boundary conditions are imposed on both parallel surfaces. The repulsive Casimir force is associated with the bulk spacetime containing a naked singularity,  which is generally considered forbidden. Additionally, we argue that topological censorship offers a natural explanation for why the vacuum of parallel defects is dual to the AdS soliton. }

\begin{document}

\maketitle

\flushbottom

\section{Introduction}

The Casimir effect \cite{Casimir:1948dh} arises from the alteration of vacuum zero-point energy due to the presence of boundaries or specific background geometries \cite{Plunien:1986ca, Bordag:2001qi, Milton:2004ya, Bordag:2009zz}. It plays a significant role in various fields, including condensed matter physics, cosmology \cite{Li:2009pm, Wang:2016och}, and high-energy physics \cite{Vepstas:1984sw}. The Casimir effect has been experimentally measured and holds potential applications in nanotechnology \cite{Mohideen:1998iz, Bressi:2002fr, Klimchitskaya:2009cw}. Interestingly, it may also support the existence of traversable wormholes \cite{Morris:1988tu, Maldacena:2020sxe}. While the Casimir effect has been widely researched, much of the existing work focuses on free fields. Investigating this effect in strongly coupled field theories could reveal new features absent in free-field systems. Recent developments in this area include studies of the Casimir effect in strongly coupled Yang-Mills theories \cite{Chernodub:2018pmt,Karabali:2025olx}, based on lattice numerical simulations or nonperturbative Hamiltonian analysis. On the other hand, gauge/gravity duality serves as a powerful framework for these studies \cite{Maldacena:1997re}. Recent developments in this area are noteworthy. First, \cite{Miao:2024ddp} constructs the gravity dual of a wedge space and finds that the displacement operator universally determines the wedge Casimir effect in the smooth limit. Second, \cite{Miao:2024gcq, Miao:2025utb} conjecture that the AdS/BCFT \cite{Takayanagi:2011zk} with minimal brane tension establishes a universal lower bound for the Casimir effect. This conjecture has been proven for general 2D  boundary conformal field theories (BCFTs) and has passed tests involving free theories, the Ising model, and \(O(N)\) models with \(N=2,3\) for 3D BCFTs. Interestingly, unlike the Kovtun-Son-Starinets (KSS) bound for fluids \cite{Kovtun:2004de}, the holographic bound for the Casimir effect is independent of the choices of gravity theories. 

We are interested in the holographic Casimir effect for two parallel defects (PD), including plane, hyperbolic, and spherical defects. The metric is given by
\begin{align}\label{Intro: BCFT metric}
ds_{\text{PD}}^2=dx^2+ R^2 \Big(d\theta^2+\frac{\sin(\sqrt{k} \theta)^2}{k}  d\Omega_{d-2}^2 \Big), \ \ \ \ 0\le x \le L.
\end{align}
Here, \( R \) represents the defect radius (with \( R = 1 \) for a plane defect), \( d\Omega_{d-2}^2 \) is the line element of a \((d-2)\)-dimensional unit sphere, \( L \) is the distance between the two defects, and \( k \in \{0, -1, 1\} \) corresponds to plane, hyperbolic, and spherical defects, respectively. For simplicity, we focus on the Euclidean signature in this paper. 
By examining the symmetry of the geometry, the energy conservation condition \( \nabla_i \langle T^{ij} \rangle = 0 \), and the Weyl anomaly \( \langle T^{i}_{\ i} \rangle = \mathcal{A} \), we derive the vacuum expectation value of the energy-momentum tensor for the parallel defects:
\begin{align}\label{Intro: Casimir effect}
\langle T^{i}_{\ j} \rangle=\frac{\kappa_1}{L^d} \text{diag}\Big(-(d-1),1,\dots,1 \Big)+ \Big( \text{anomalous terms} \sim \frac{A}{R^d} \Big),
\end{align}
where \( \kappa_1 \) denotes the Casimir amplitude, a dimensionless constant for plane defects, and a dimensionless function of \( L/R \) for hyperbolic and spherical defects. The anomalous terms vanish for \( k = 0 \) and depend only on the A-type central charge $A$ and the radius \( R \) for \( k \in \{-1, 1\} \), which are universally determined by the Weyl anomaly. Since these terms are independent of the width \( L \), we do not provide their exact expressions in this paper. For more details on these anomalous terms, please refer to \cite{Herzog:2013ed}.
In the limit of large radius \( L/R \to 0 \), the hyperbolic and spherical defects converge to the plane defects. Therefore, we conclude:
\begin{align}\label{Intro: large R limit}
\lim_{R\to \infty} \kappa_1|_{k\in \{-1, 1\} }=  \kappa_1|_{k=0}. 
\end{align}

Note that the metric (\ref{Intro: BCFT metric}) is conformally flat. The parallel hyperbolic defects are conformally equivalent to a wedge space \cite{Miao:2024ddp}. To demonstrate this, we can substitute the hyperbolic metric in the large parentheses of (\ref{Intro: BCFT metric}) with the Poincaré AdS metric. Consequently, the metric (\ref{Intro: BCFT metric}) with \( k=-1 \) becomes clearly conformally equivalent to a wedge metric:
\begin{align}\label{Intro: hyperbolic wedge}
ds_{\text{PD}}^2|_{k=-1}=dx^2+ R^2\Big( \frac{dr^2+dy_a^2}{r^2} \Big)=\frac{R^2}{r^2}\Big( r^2 d\phi^2+ dr^2 +dy_a^2\Big), \ \ 0\le \phi \le L/R,
\end{align}
where $\phi=x/R$. Similarly, the spherical defects are conformally equivalent to a cylindrical annulus. To illustrate this, we perform the coordinate transformation \( dx = R \, dr/r\) and rewrite the metric (\ref{Intro: BCFT metric}) with \( k=1 \) as follows:
\begin{align}\label{Intro: spherical annulus}
ds_{\text{PD}}^2|_{k=1}=\frac{R^2}{r^2}\Big( dr^2+ r^2 d\Omega_{d-1}^2 \Big), \ \ R \le r \le R e^{L/R}.
\end{align}
By applying the Weyl transformations (\ref{Intro: hyperbolic wedge},\ref{Intro: spherical annulus}) for (\ref{Intro: Casimir effect}), we obtain the Casimir stress tensors for the wedge and cylindrical annulus:
\begin{align}\label{Intro: Casimir effect flat}
 \langle T^{i}_{\ j} \rangle_{\text{flat}} = \frac{\kappa_2}{r^d} \text{diag}\Big(-(d-1), 1, \ldots, 1 \Big),
\end{align} 
where \(\kappa_2=(R/L)^d \kappa_1 \) is a dimensionless function of \(L/R\), and the anomalous terms vanish in flat space. The region far from the corner of the wedge approaches that of a parallel plane in the limit \(r \to \infty\) and \(r \Omega \to L\), where \(\Omega = L/R\) is the opening angle of the wedge \cite{Deutsch:1978sc}. Similarly, the cylindrical annulus approaches a parallel plane in the limit \(R \to \infty\) and \(L/R \to 0\). These observations lead to the following limit for the Casimir amplitude of the wedge and cylindrical annulus: 
\begin{align}\label{Intro: large R limit flat}
 \lim_{R \to \infty} \kappa_2 \to \frac{\kappa_1|_{k=0}}{\left(\frac{L}{R}\right)^d}, 
 \end{align} 
which agrees with (\ref{Intro: large R limit}) for $\kappa_2= (R/L)^d\kappa_1$. 

It has been conjectured that the negative Casimir amplitude \(-\kappa_1\) relative to the norm of the displacement operator \(C_D\) has a lower bound for fixed system sizes \(L\) and \(R\) \cite{Miao:2024gcq, Miao:2025utb}:
\begin{align}\label{Intro: Casimir bound}
-\frac{\kappa_1}{C_D} \ge \lim_{T\to T_{\text{min}}} \left(-\frac{\kappa_1}{C_D}\right)|_{\text{holo}}.
\end{align} 
Here, the lower bound is given by the AdS/BCFT with the minimal brane tension $T_{\text{min}}=-(d-1)$. It means that, for fixed geometry, the Casimir energy density $-\kappa_1$ per degree of freedom $C_D$ has a fundamental lower bound determined by the holographic CFTs with the minimal boundary degrees of freedom $T_{\text{min}}$. Additionally, it is important to note that \(\kappa_1\) can be substituted with \(\kappa_2 = (R/L)^d \kappa_1\) when considering fixed system sizes \(L\) and \(R\). \cite{Miao:2024gcq, Miao:2025utb} have investigated the holographic bounds of the Casimir effect for parallel planes and hyperbolic defects. In this paper, we provide evidence that the bound (\ref{Intro: Casimir bound}) also applies to parallel spherical defects. We compare results from free and holographic theories, confirming that holographic theories with minimal brane tension yield a smaller Casimir effect for parallel spherical defects.

The gravity duals of parallel defects are represented by a portion of the AdS soliton \cite{Takayanagi:2011zk, Miao:2024ddp}:
\begin{align}\label{Intro: holo defect}
ds^2=\frac{\frac{dz^2}{h(z)}+(h(z)-1)dx^2+ds_{\text{PD}}^2}{z^2}, \ \ h(z)=1+k \frac{z^2}{R^2}-c_1 z^d,
\end{align}
where we have set the AdS radius to be one, \( ds_{\text{PD}}^2 \) is defined in (\ref{Intro: BCFT metric}), and \( k = (0, -1, 1) \) corresponds to plane, hyperbolic, and spherical defects, respectively. The constant \( c_1 \) is associated with the width \( L \). The gravity duals of parallel plane and hyperbolic defects have been investigated in \cite{Takayanagi:2011zk} and \cite{Miao:2024ddp}, respectively. In this paper, we focus on holographic spherical defects. Notably, unlike plane and hyperbolic defects, holographic spherical defects undergo a phase transition as the width \( L \) increases. A small width \( L \) corresponds to the connected phase, which is characterized by the AdS soliton (\ref{Intro: holo defect}) with a connected end-of-the-world (EOW) brane. In contrast, as \( L \) becomes large, the disconnected phase is reached, which is dual to a portion of AdS space with disconnected EOW branes. 
See Fig.~\ref{brane shapes for two c1} for a visual representation of the geometry. 
Remarkably, the Casimir effect vanishes in the disconnected phase. It is noteworthy that free theories do not exhibit such a phase transition or this vanishing at finite width, highlighting key differences in the Casimir effect between free and strongly coupled theories.

Furthermore, we observe novel relationships between the censorship principle and the holographic Casimir effect. We find that cosmic censorship prevents the divergent Casimir effect and provides a holographic explanation for the attractive Casimir force. The repulsive Casimir force is dual to a spacetime with a naked singularity, which should be ruled out. Additionally, topological censorship offers a natural resolution to the discrepancy concerning the AdS soliton: while the bulk exhibits periodicity, the AdS boundary does not.

The paper is organized as follows: In Section 2, we examine the gravity dual of parallel spherical defects and derive the holographic lower bound for the corresponding Casimir effect. Section 3 discusses the holographic phase transition associated with parallel spherical defects. As the width \(L\) increases, the gravity dual transitions from a connected phase to a disconnected phase, revealing a significant difference in the Casimir effect between free and strongly coupled theories. 
In Section 4, we examine the constraints on the Casimir effect imposed by the cosmic and topological censorships. Section 5 compares the Casimir effects in free theories and holographic theories. Notably, we confirm that the AdS/BCFT with minimal brane tension produces a smaller ratio of \((- \kappa_1 / C_D)\) compared to free theories, and we observe that there is no phase transition in free theories. Finally, we conclude with a discussion of open issues in Section 6. 
Appendices A, B, and C analyze the Casimir effect of parallel spherical defects on free scalar fields, Maxwell fields, and free Dirac fermions, respectively.

\section{Holographic Casimir effect}

This section examines the gravitational dual of parallel spherical defects and tests the holographic bound (\ref{Intro: Casimir bound}) for the Casimir effect. For simplicity, we focus on the scenario where the width \( L \) is not excessively large. In this case, the gravity dual is represented by a portion of the AdS soliton. We will address cases with large \( L \) in Section 3, when we discuss the holographic phase transition.

 \begin{figure}[htbp]
  \centering
\includegraphics[width=0.6\textwidth]{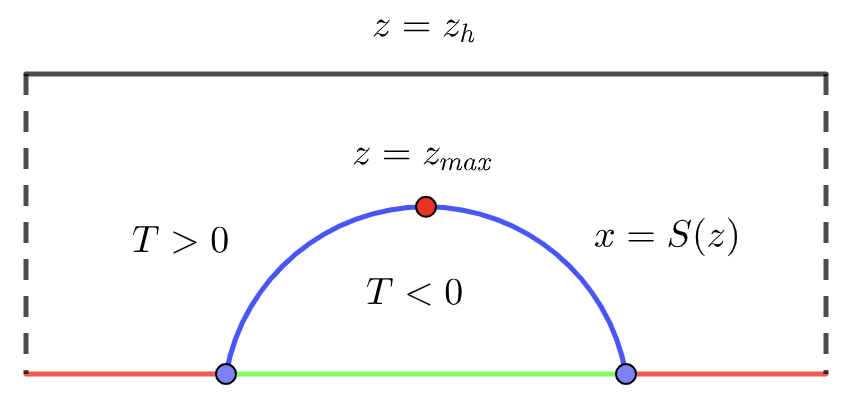}
 \caption{Gravity dual of parallel spherical defects. The blue points represent the two parallel spherical defects, while the blue curve indicates the EOW brane, i.e., $x=S(z)$. Due to the periodic nature of the bulk spacetime, the two dashed lines are identified. The gravity dual corresponds to the region between the blue curve and the green line for negative brane tension (\(T < 0\)). Conversely, the complement of this region represents the gravity dual for positive brane tension (\(T > 0\)).} 
 \label{GravityDual}
\end{figure}

\subsection{General discussion}

Let us start with the geometry of holographic parallel spherical defects. See Fig. \ref{GravityDual}. The blue points represent the two parallel spherical defects, while the blue curve indicates the EOW brane. Due to the periodic nature of the bulk spacetime, the two dashed lines are identified. The gravity dual corresponds to the region between the blue curve and the green line for negative brane tension (\(T < 0\)). Conversely, the complement of this region represents the gravity dual for positive brane tension (\(T > 0\)). 

The metric of the AdS soliton and the embedding function of the EOW brane are given by
\begin{align}\label{sect 2: holo sphere defect}
&\text{metric}:\ ds^2=\frac{\frac{dz^2}{h(z)}+h(z)dx^2+R^2 d\Omega_{d-1}^2}{z^2}, \ \ h(z)=1+\frac{z^2}{R^2}-c_1 z^d,\\
&\text{brane}: \ \ x=S(z), \label{sect 2: Sz}
\end{align}
where \( R \) is the radius of the sphere, \( d\Omega_{d-1}^2 \) is the line element of the unit sphere in \((d-1)\) dimensions, and \( S(z) \) is the embedding function of the brane. We can parametrize \( c_1 \) as:
\begin{align}\label{sect 2: c1}
c_1=\frac{1}{z_h^d}(1+\frac{z_h^2}{R^2}),
\end{align}
so that $h(z_h)=0$. To eliminate the conical singularity in the bulk, we derive the bulk period of \( x \) as:
\begin{align}\label{sect 2: beta}
\beta=\frac{4\pi}{|h'(z_h)|}=\frac{4 \pi  R^2 z_h}{(d-2) z_h^2+d R^2}.
\end{align}
On the AdS boundary, the range of \( x \) is labeled as \( 0 \le x \le L \). For a positive brane tension \( T \ge 0 \), the bulk dual includes the ``horizon'' at \( z = z_h \). Therefore, we require \( L \le \beta \) to avoid the conical singularity \cite{Fujita:2011fp, Miyaji:2021ktr}. Fortunately, this constraint is typically satisfied automatically \cite{Miyaji:2021ktr}. On the other hand, for negative brane tension $T\le 0$, the potential conical singularity is hidden behind the EOW brane \cite{Miao:2025utb}. As a result, the conical singularity is irrelevant, and $L$ can exceed $\beta$ for negative brane tension \cite{Miao:2025utb}. By applying the holographic renormalization \cite{deHaro:2000vlm}, we derive the pressure
\begin{align}\label{sect 2: Txx}
T_{xx}=-(d-1)c_1+ \text{anomalous terms}. 
\end{align} 
Comparing this expression with (\ref{Intro: Casimir effect}), we obtain the holographic Casimir amplitude
\begin{eqnarray}\label{sect2: kappa1}
\kappa_1=c_1 L^d=\frac{L^d}{z_h^d}(1+\frac{z_h^2}{R^2}).
\end{eqnarray}

We impose Neumann boundary condition (NBC) on the EOW brane \cite{Takayanagi:2011zk} 
\begin{eqnarray}\label{sect 2: NBC}
K_{ij}-(K-T) h_{ij}=0,
\end{eqnarray}
where \( K_{ij} = h_i^k h_j^l \nabla_k n_l \) represents the extrinsic curvature, \( n_l \) is the outward-pointing normal vector, \( h_{ij} \) is the induced metric, and \( T = (d - 1) \tanh(\rho) \) is the brane tension. Note that since $n_l$ is outward-pointing, $K_{ij}$ flips sign when crossing the EOW brane. Equivalently, the tension $T$ in the NBC (\ref{sect 2: NBC}) changes sign when crossing the EOW brane. It implies that the gravity dual for \( T > 0 \) is the complement of the gravity dual for \( T < 0 \). Let us first examine the case of \( T < 0 \). By substituting (\ref{sect 2: holo sphere defect},\ref{sect 2: Sz}) into the NBC (\ref{sect 2: NBC}), we obtain:
\begin{equation}\label{sect 2: dS}
	S '(z) =\frac{-T}{h(z) \sqrt{(d-1)^2 h(z) - T^2}}.
\end{equation}
Here, we focus on the left half of the blue curve in Fig. \ref{GravityDual} for \( T < 0 \). The width \( L \) for negative brane tension is given by:
\begin{equation}\label{sect 2: Ln}
	L_{n}(T)=2\int_0^{z_{\text{max}}}S'(z) dz=\int_0^{z_{\text{max}}}  \frac{-2T}{h(z) \sqrt{(d-1)^2 h(z) - T^2}}dz, \ \ \text{for}\ T\le 0,
\end{equation}
where \( z_{\text{max}} \) is the turning point that satisfies \( (d - 1)^2 h(z_{\text{max}}) - T^2 = 0 \). It is illustrated by the red point in Fig. \ref{GravityDual}. Since the gravity dual for \( T > 0 \) is the complement of the gravity dual for \( T < 0 \), we can express the width \( L \) for positive brane tension as follows:
\begin{equation}\label{sect 2: Lp}
	L_{p}(T)=\beta-L_n(T\to -T) , \ \ \text{for}\ T\ge 0,
\end{equation}
where \( \beta \) represents the bulk period defined in (\ref{sect 2: beta}).

From (\ref{sect2: kappa1},\ref{sect 2: Ln},\ref{sect 2: Lp}), we can derive the holographic Casimir amplitude $\kappa_1$. To investigate the holographic bound of $-\kappa_1/C_D$(\ref{Intro: Casimir bound}), we also need information about the displacement operator. 
For general brane tension \( T = (d - 1) \tanh(\rho)  \) , $C_D$ is given by \cite{Miao:2024ddp}: 
\begin{eqnarray}\label{sect2: CD}
C_D=\frac{2 (d-1) \Gamma (d+2)}{ \pi ^{\frac{d-2}{2}}}
\begin{cases}
\frac{-\text{sech}^3\left(\rho \right) \left(-\text{csch}\left(\rho \right)\right){}^{-d}}{\sqrt{\pi } \Gamma \left(\frac{d+1}{2}\right) \left(\text{sech}^3\left(\rho \right) F+d \text{csch}\left(\rho \right) \left(-\tanh \left(\rho \right)\right){}^d\right)}, \ \ \ \ \ \ \ \ \ \ \ \ \ \ \ \ \ \ \text{for} \rho\le 0,\\
\frac{1}{\pi  (d-1) d \Gamma \left(\frac{d}{2}\right)-\frac{\sqrt{\pi } \Gamma \left(\frac{d+1}{2}\right) \left(d \left(\tanh \left(\rho \right) \text{sech}\left(\rho \right)\right){}^d-\sinh \left(\rho \right) \text{sech}^{d+3}\left(\rho \right) F\right)}{\text{sech}^2\left(\rho \right) \tanh ^{d+1}\left(\rho \right)}},\text{for } \rho\ge 0,
\end{cases}
\end{eqnarray} 
where $F=\, _2F_1\left(\frac{d-1}{2},\frac{d}{2};\frac{d+2}{2};-\text{csch}^2\left(\rho \right)\right)$ is the hypergeometric function. 
For dimensions \(d=3\) and \(d=4\), the expressions for \(C_D\) can be simplified as:
\begin{align}\label{sect2: CD 3d 4d}
C_D=\begin{cases}
 \frac{32}{\pi  \left(2 \tan ^{-1}\left(\tanh \left(\frac{\rho }{2}\right)\right)+\frac{\pi }{2}\right)},&
\text{for} \ d=3,\\
  \frac{120 e^{-\rho } \cosh (\rho )}{\pi ^2},&
\text{for} \ d=4.
\end{cases}
\end{align}

 \begin{figure}[htbp]
  \centering
\includegraphics[width=0.6\textwidth]{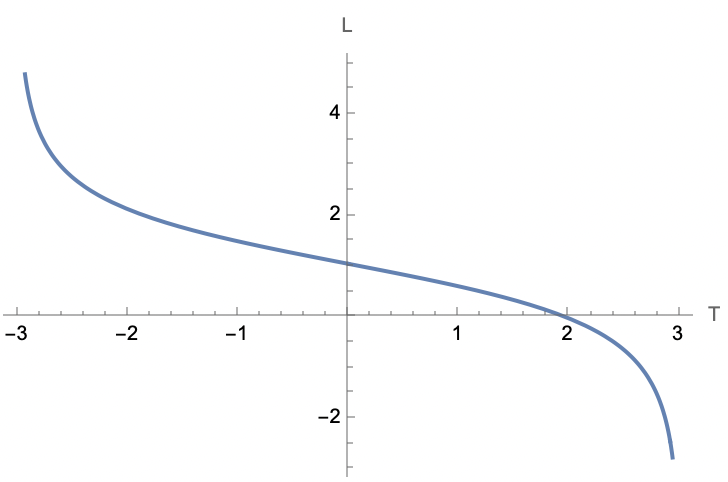}
 \caption{The width $L$ decreases with the tension $T$. Without loss of generality, we choose the parameter $d=4, z_h=1, R=1$. It shows $L$ goes to infinity as $T\to -3$, and $L$ reduces to zero at $T\approx 1.97$. The requirement $L\ge 0$ imposes an upper bound on the tension $T\le 1.97$. } 
 \label{relation LT}
\end{figure}

Before studying the ratio $-\kappa_1/C_D$, let's discuss the width $L$ between the two spherical defects.

\begin{figure}[htbp]
  \centering
\includegraphics[width=0.6\textwidth]{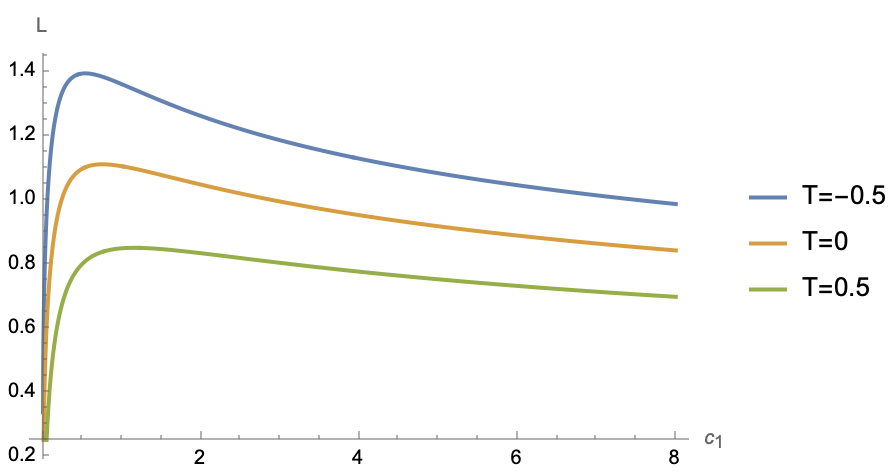}
 \caption{The width $L$ has a maximum for fixed $R$ and $T>-(d-1)$. Without loss of generality, we choose the parameter $d=4, R=1$ and $T=(-0.5,0,0.5)$. The figure shows that one $L$ corresponds to two $c_1$ (\ref{sect 2: c1}) generally. We choose the larger $c_1$ since it gives smaller free energy. Note that the smaller the brane tension, the larger the maximum of $L$. As we will discuss below, there is no upper bound of $L$ for the minimal brane tension $T\to -(d-1)$. } 
 \label{relation Lc1}
\end{figure}

\begin{itemize}

 \item{For the tensionless case \( T = 0 \), we derive an analytical expression for the width from (\ref{sect 2: Lp})}
 \begin{align}\label{sect 2: L0}
	L=L_{p}(0)=L_n(0)=\frac{\beta}{2}=\frac{2 \pi  R^2 z_h}{(d-2) z_h^2+d R^2}.
\end{align} 

  \item{The requirement \( L \ge 0 \) establishes an upper bound on the brane tension \( T \).}
  
 As discussed earlier, \( L_n \) can exceed \( \beta \) for negative values of \( T \), resulting in a negative width \( L_p < 0 \) for the corresponding positive \( T \) from (\ref{sect 2: Lp}). Thus, the condition \( L_p \ge 0 \) imposes an upper limit on the brane tension \( T \). For example, with \( d = 4 \), \( z_h = 1 \), and \( R = 1 \), as illustrated in Fig. \ref{relation LT}, the width \( L \) decreases as the tension \( T \) increases, approaching zero when \( T \approx 1.97 \). Therefore, we find that \( -3 \le T \le 1.97 \) for \( d = 4 \), \( z_h = 1 \), and \( R = 1 \).
   
    \item{For \( T > -(d-1) \), the width \( L \) is bounded above for a fixed radius \( R \). It indicates that the AdS soliton can only serve as the gravitational dual of parallel spherical defects for widths \( L \) that are not excessively large. It also suggests that the gravitational solution may undergo a phase transition as \( L \) increases, which we will discuss in detail in Section 3.}
    
  Let us first analyze the tensionless case. From (\ref{sect 2: L0}) and using the condition \( dL/dz_h= 0 \), we can determine the extremum point at \( z_h = \frac{R\sqrt{d}}{\sqrt{d-2}} \), yielding the maximum width:
     \begin{align}\label{sect 2: L0 maximum}
	\frac{L}{R}|_{T=0}\le \frac{\pi }{\sqrt{(d-2) d}}.
	\end{align} 
For nonzero brane tension, Fig. \ref{relation Lc1} suggests that \( L/R \) also typically has a maximum value. Additionally, this maximum width \( L \) decreases as the brane tension increases. Fig. \ref{relation Lc1} shows that one $L$ corresponds to two $c_1$ (\ref{sect 2: c1}) generally. We choose the larger $c_1$ since it gives smaller free energy. 

 \item{For minimal brane tension \( T \to -(d-1) \), the width \( L \) is unbounded. }
    
Fig. \ref{relation LT}  indicates that for a fixed value of \(c_1\), \(L\) approaches infinity as \(T\) approaches \(-(d-1)\). Similarly, Fig. \ref{relation Lc1} demonstrates that for a fixed \(T\) with varying \(c_1\), the maximum value of \(L\) increases as \(T\) decreases. It also implies $L\to \infty$ as $T\to -(d-1)$. In the limit where \(c_1\) approaches zero, we find \(z_{\text{max}}\) goes to infinity, as indicated by the turning-point condition \(h(z_{\text{max}}) - \tanh(\rho)^2 = 0\). Consequently, the width (\ref{sect 2: Ln}), for \(c_1 \to 0\) becomes 
 \begin{eqnarray}\label{sect2: L zero c1}
 L|_{c_1 \to 0}=  \int_0^{\infty } \frac{-2 \tanh (\rho )}{\left(z^2+1\right) \sqrt{1+z^2-\tanh ^2(\rho )}} \, dz=-2\rho,
\end{eqnarray}
which confirms that $L$ is unbounded in the minimal tension limit as \(\rho \to -\infty\). 
 \end{itemize}

\subsection{Holographic lower bound}

We are now prepared to examine the lower bound of the ratio \(-\kappa_1/C_D\), which is proposed to be given by the AdS/BCFT with minimal brane tension \(T \to -(d-1)\) or \(\rho \to -\infty\). For simplicity, we set the sphere radius $R=1$ below. 

In the limit of minimal tension, where \(x \equiv \mathrm{sech}^2\rho \to 0\),  $C_D$ behaves as
\begin{equation}\label{sect 2: CD limit}
	 C_D = \frac{2\pi^{\frac{1-d}{2}} (d-1) \Gamma(d+2)}{d \, \Gamma\!\left(\frac{d+1}{2}\right)} \, x^{1-\frac{d}{2}} + \mathcal{O}\!\left(x^{2-d/2}\right).
\end{equation}
We make the following rescales
\begin{equation}\label{sect 2: c1 zmax limit}
c_1 = \hat{c}_1 \, x^{1-\frac{d}{2}}, \qquad z_{\mathrm{max}} = \hat{z}_{\mathrm{max}} \sqrt{x},
\end{equation}
to ensure that we obtain a finite ratio:
\begin{equation}\label{sect 2: ratio limit}
\lim_{x\to 0} -\frac{\kappa_1}{C_D}=-\frac{c_1 L^d}{C_D}=-\frac{ d \Gamma \left(\frac{d+1}{2}\right)}{2 \pi ^{\frac{1-d}{2}} (d-1) \Gamma (d+2)} \hat{c}_1 L^d,
\end{equation}
where \(L\) remains finite. In the rescaled variables, the turning-point condition given by \((d - 1)^2 h(z_{\text{max}}) - T^2 = 0\) transforms into:
\begin{equation}\label{sect 2: c1 zmax relation}
1 + \hat{z}_{\mathrm{max}}^2 - \hat{c}_1 \hat{z}_{\mathrm{max}}^{\,d} = 0,
\end{equation}
where we have set $R=1$. By changing variables to \(y = z / z_{\mathrm{max}}\) and taking the limit as \(x \to 0\), we can derive a numerically manageable expression for \(L\) from (\ref{sect 2: Ln}):
\begin{equation} \label{sect 2: L small T}
	L = \int_0^1 dy  \frac{2 \hat{z}_{\text{max}}}{\sqrt{1 + \hat{z}_{\text{max}}^2 y^2 - (1 + \hat{z}_{\text{max}}^2) y^{d}}}.
\end{equation}
Note that $L \to \infty$ as $\hat{z}_{\text{max}}\to \infty$, which means there is no upper bound for the width $L$ for the minimal brane tension $T=-(d-1)$. 

\begin{figure}[htbp]
  \centering
\includegraphics[width=0.4\textwidth]{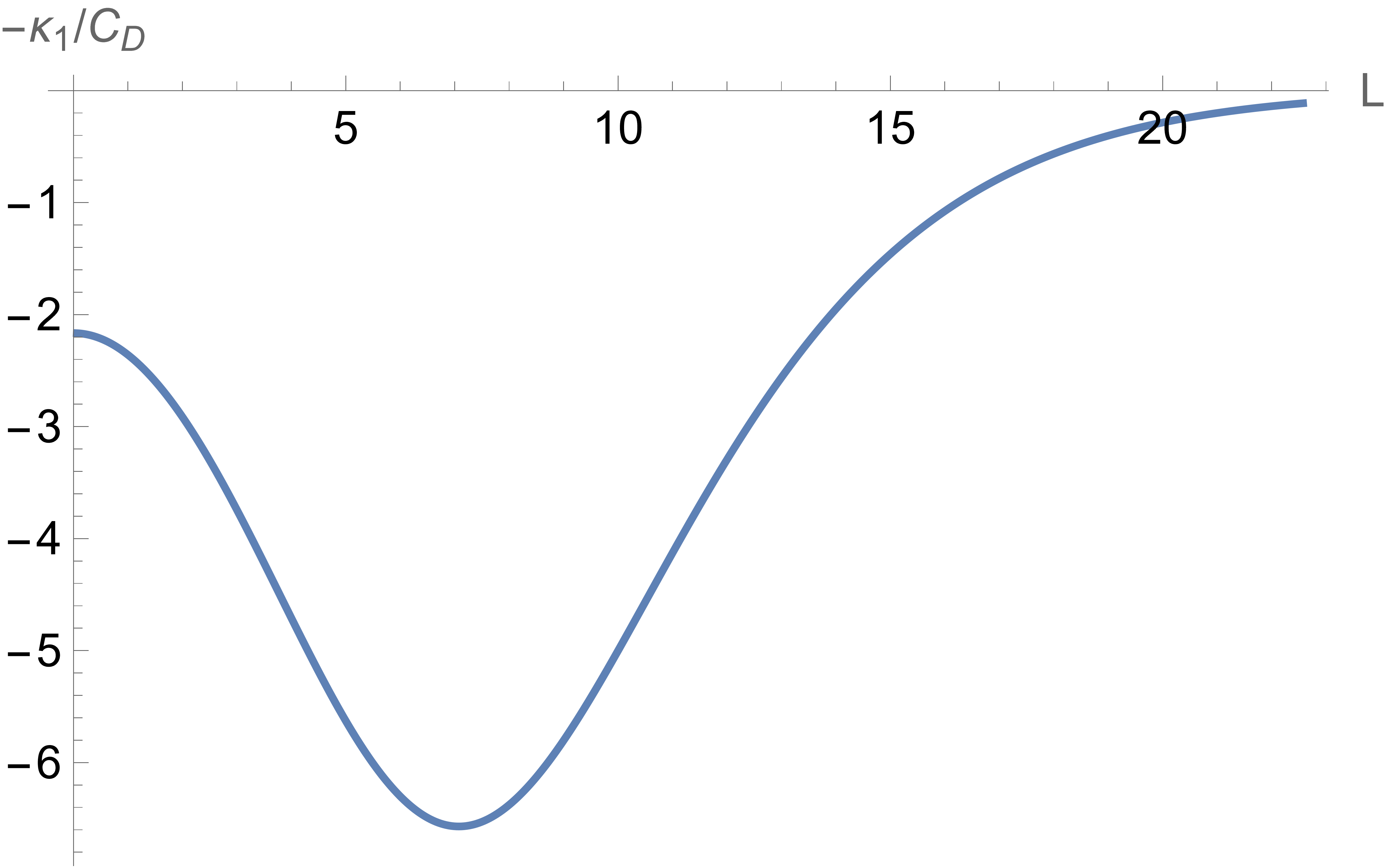} \includegraphics[width=0.4\textwidth]{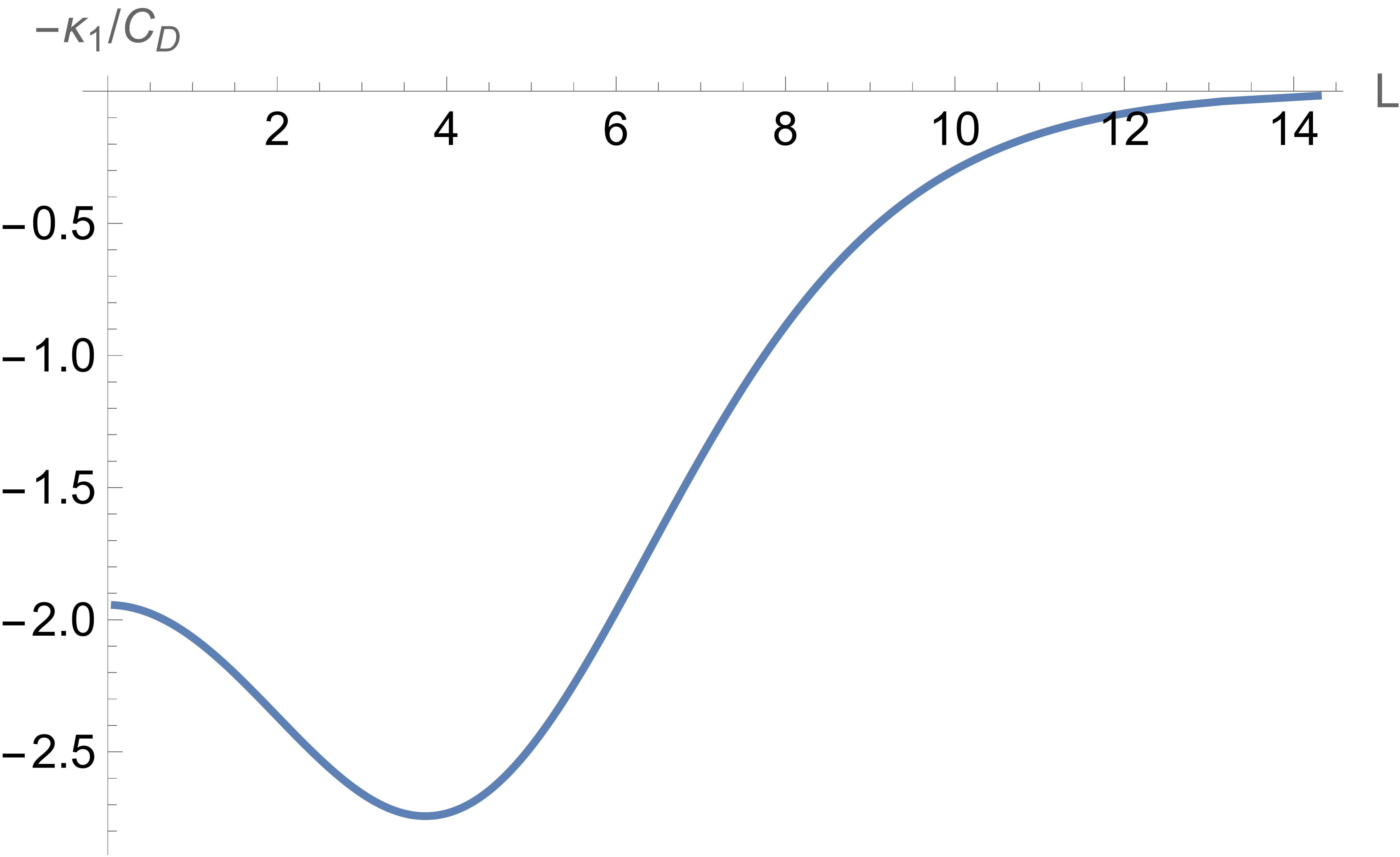}
 \caption{Holographic ratio $-\kappa_1/C_D$ versus width $L$ in the minimal-tension limit for $d=3$ (left) and $d=4$ (right). As \( L \) approaches 0, the curvature of the sphere becomes negligible, and the ratio simplifies to that of parallel plane defects: $-2.17$ for $d=3$ and $-1.94$ for $d=4$. Conversely, as \( L \) approaches infinity, the ratio tends toward zero. Notably, the figure indicates that the ratio has a minimum value. } 
 \label{3d4d ratio}
\end{figure}

For a given value of \(\hat{z}_{\text{max}}\), we can determine \(\hat{c}_1\) and \(L\) from \((\ref{sect 2: c1 zmax relation})\) and \((\ref{sect 2: L small T})\), and then calculate the ratio \((\ref{sect 2: ratio limit})\). Fig. \ref{3d4d ratio} displays the resulting ratio $-\kappa_1/C_D$ as a function of $L$ for $d=3,4$. As \( L \) approaches 0, the curvature of the sphere becomes negligible, and the ratio simplifies to that of parallel plane defects \cite{Miao:2025utb}: $-2.17$ for $d=3$ and $-1.94$ for $d=4$. Conversely, as \( L \) approaches infinity, the ratio tends toward zero. Notably, Fig. \ref{3d4d ratio} indicates that the ratio has a minimum value.

\begin{figure}[htbp]
	\centering
	\begin{subfigure}{0.48\textwidth}
		\centering
		\includegraphics[width=\textwidth]{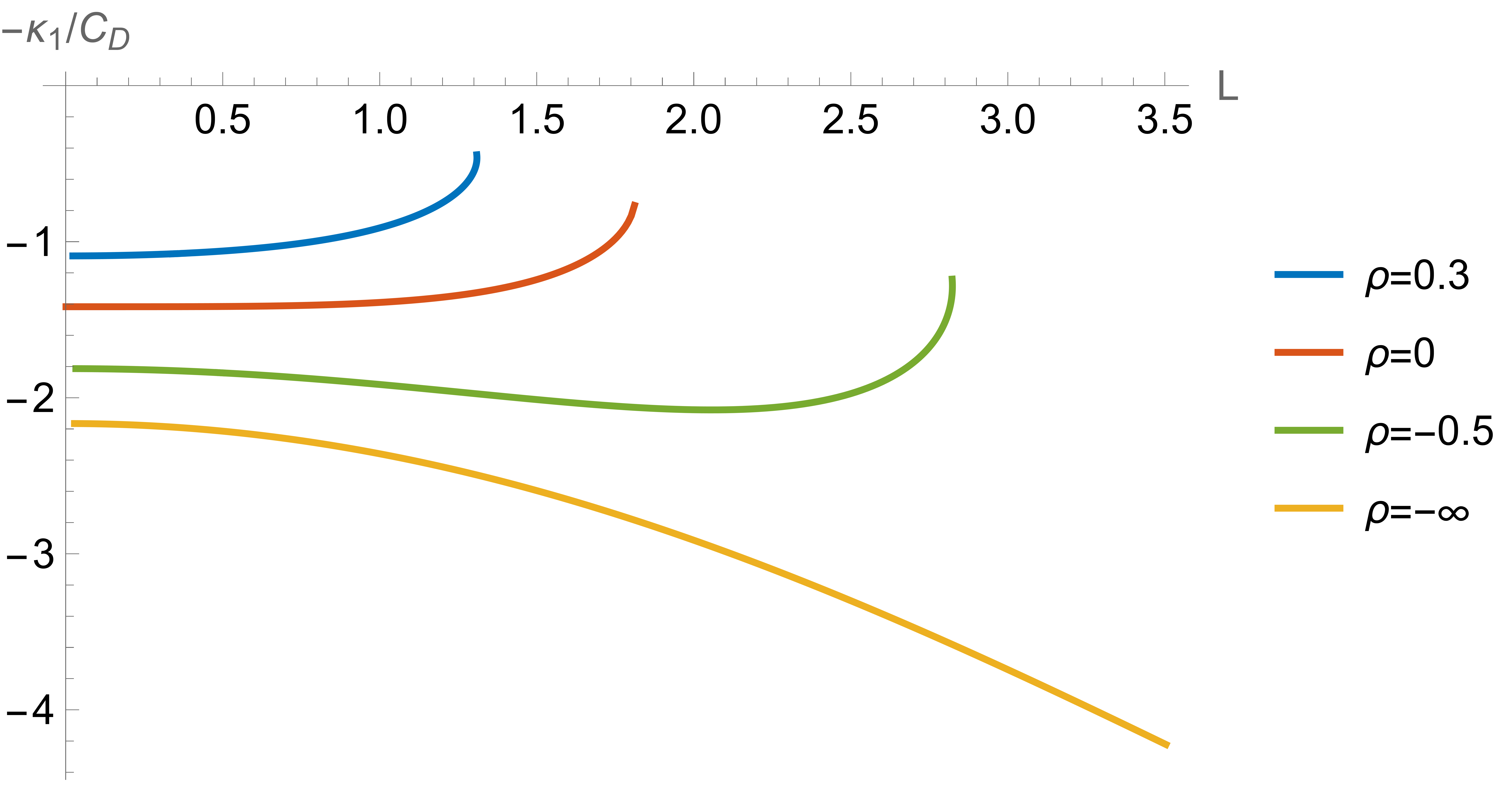}
		\caption{$d=3$}
	\end{subfigure}
	\hfill
	\begin{subfigure}{0.48\textwidth}
		\centering
		\includegraphics[width=\textwidth]{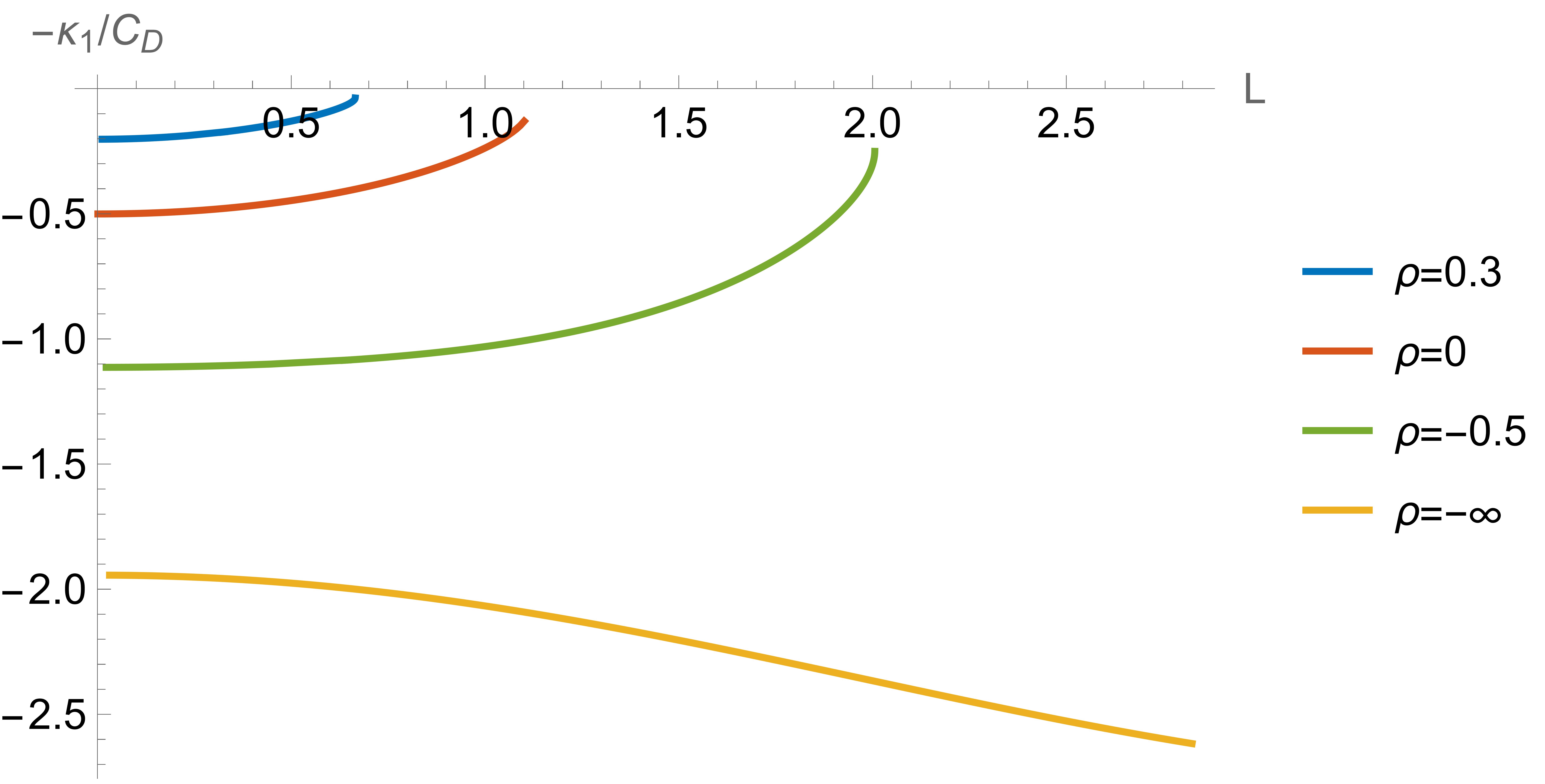}
		\caption{$d=4$}
	\end{subfigure}
	\caption{Ratio $-\kappa_1/C_D$ for various tensions for $d=3$ (left) and $d=4$ (right). The minimal tension \( T \to -(d-1) \) results in the least ratio of \( -\kappa_1/C_D \). Moreover, for tensions where \( T > -(d-1) \), there is an upper limit on the width \( L \). If the width exceeds this maximum value, the gravitational dual transitions from an AdS soliton to an AdS space, and in this scenario, \( \kappa_1 = 0 \). } 
\label{ratio for general T}
\end{figure}

Let us go on to discuss the holographic ratio $-\kappa_1/C_D$ for general brane tensions. From (\ref{sect2: kappa1},\ref{sect 2: Ln},\ref{sect 2: Lp}) for $\kappa_1$ and (\ref{sect2: CD 3d 4d}) for $C_D$, we can derive numerically the ratio $-\kappa_1/C_D$. As shown in Fig. \ref{ratio for general T}, the minimal tension \( T \to -(d-1) \) results in the lowest ratio of \( -\kappa_1/C_D \). Additionally, when the tension \( T \) exceeds \( -(d-1) \), there is an upper limit on the width \( L \). If \( L \) surpasses this maximum value, the gravitational dual transitions from an AdS soliton to an AdS space, and in this scenario, \( \kappa_1 = 0 \). We will discuss this point further in Section~3.

In summary, we have examined the gravity dual of parallel spherical defects in this section. This dual is represented by a portion of the AdS soliton when the width \( L \) is less than the maximum width \( L_{\max} \) and the tension \( T \) is greater than \(- (d-1)\). Refer to Fig. \ref{relation Lc1} for examples of \( L_{\max} \). It is important to note that there is no upper limit on the width for the minimal brane tension, meaning that as \( T \) approaches \(- (d-1)\), \( L_{\max} \) tends to infinity. Additionally, we demonstrate that the AdS/BCFT with minimal brane tension establishes a holographic lower bound on the Casimir effect, as illustrated in Fig. \ref{ratio for general T}. We will perform more tests of
this lower bound using free theories in Section 5.

\section{Holographic phase transition}

In this section, we investigate the holographic phase transition of the Casimir effect concerning parallel spherical defects. In the previous section, we observed a puzzling feature of the AdS soliton solution with \(c_1 > 0\): the boundary width \(L\) cannot exceed a maximum value, \(L_{\max}\). This is illustrated in Fig. \ref{relation Lc1}. There is no such constraint for field theories, which suggests that as the boundary width varies, the same BCFT may be associated with more than one classical bulk saddle. The dominant saddle corresponds to the one with the lower free energy or, equivalently, a smaller renormalized Euclidean action.

We propose that the second saddle is represented by AdS space with disconnected end-of-the-world (EOW) branes, which we refer to as the disconnected phase. Conversely, we call the AdS soliton with connected EOW branes the connected phase. In the disconnected phase, we have \(c_1 = 0\), whereas in the connected phase, \(c_1 > 0\). 
The case of \(c_1 < 0\) is unphysical and will be discussed further in Section 4.
As we increase the boundary width \(L\), a first-order phase transition occurs, resulting in a discontinuous pressure. Notably, the pressure vanishes in the disconnected phase, indicating that there is no Casimir effect for sufficiently large widths. This behavior is quite distinct from that observed in free theories and is a unique characteristic of the Casimir effect in strongly coupled field theories with a gravity dual.

\subsection{Disconnected phase}

\begin{figure}[htbp]
	\centering
	\begin{subfigure}{0.45\textwidth}
		\centering
		\includegraphics[width=\textwidth]{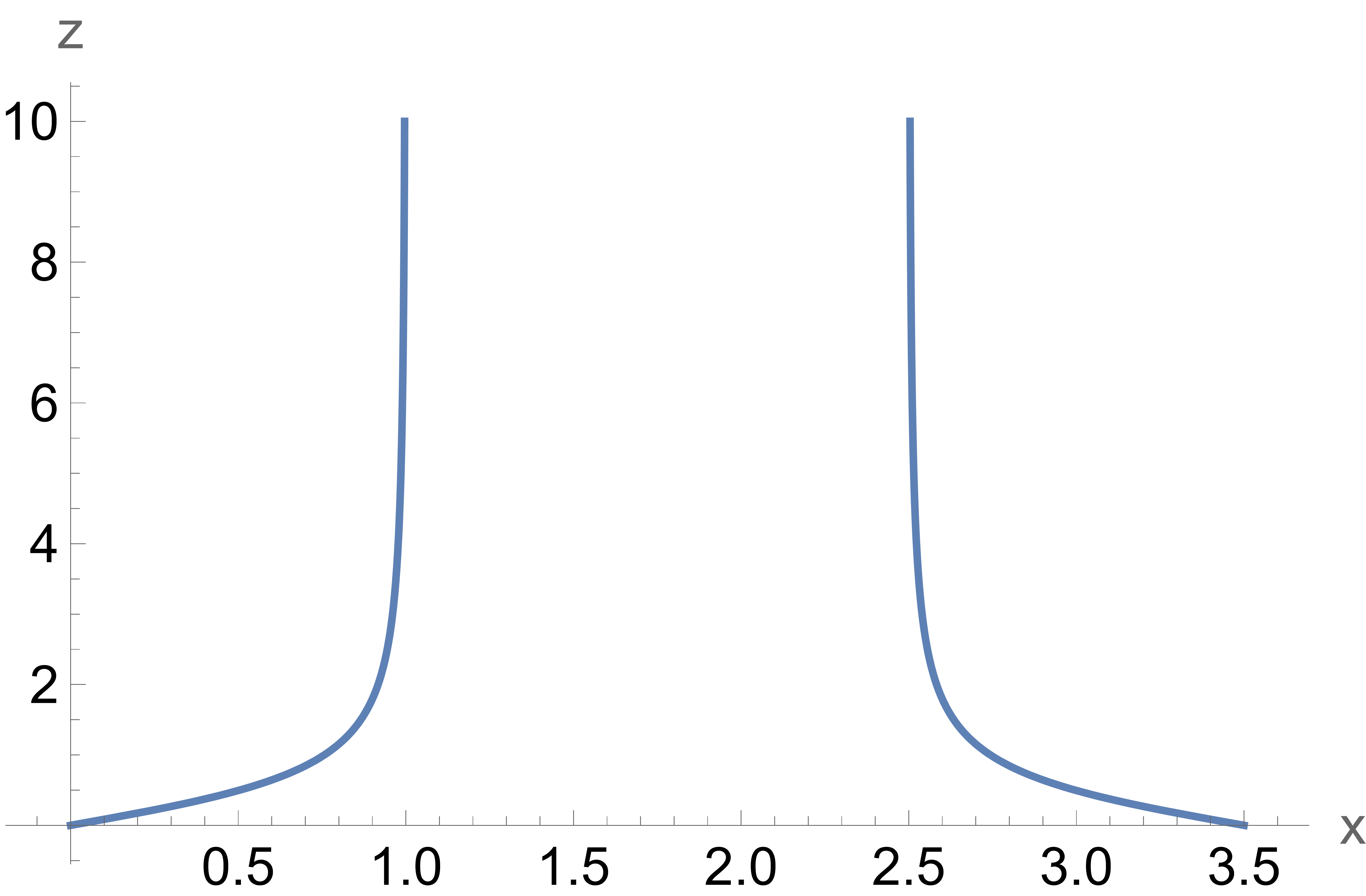}
		\caption{$c_1=0$}
	\end{subfigure}
	\hfill
	\begin{subfigure}{0.45\textwidth}
		\centering
		\includegraphics[width=\textwidth]{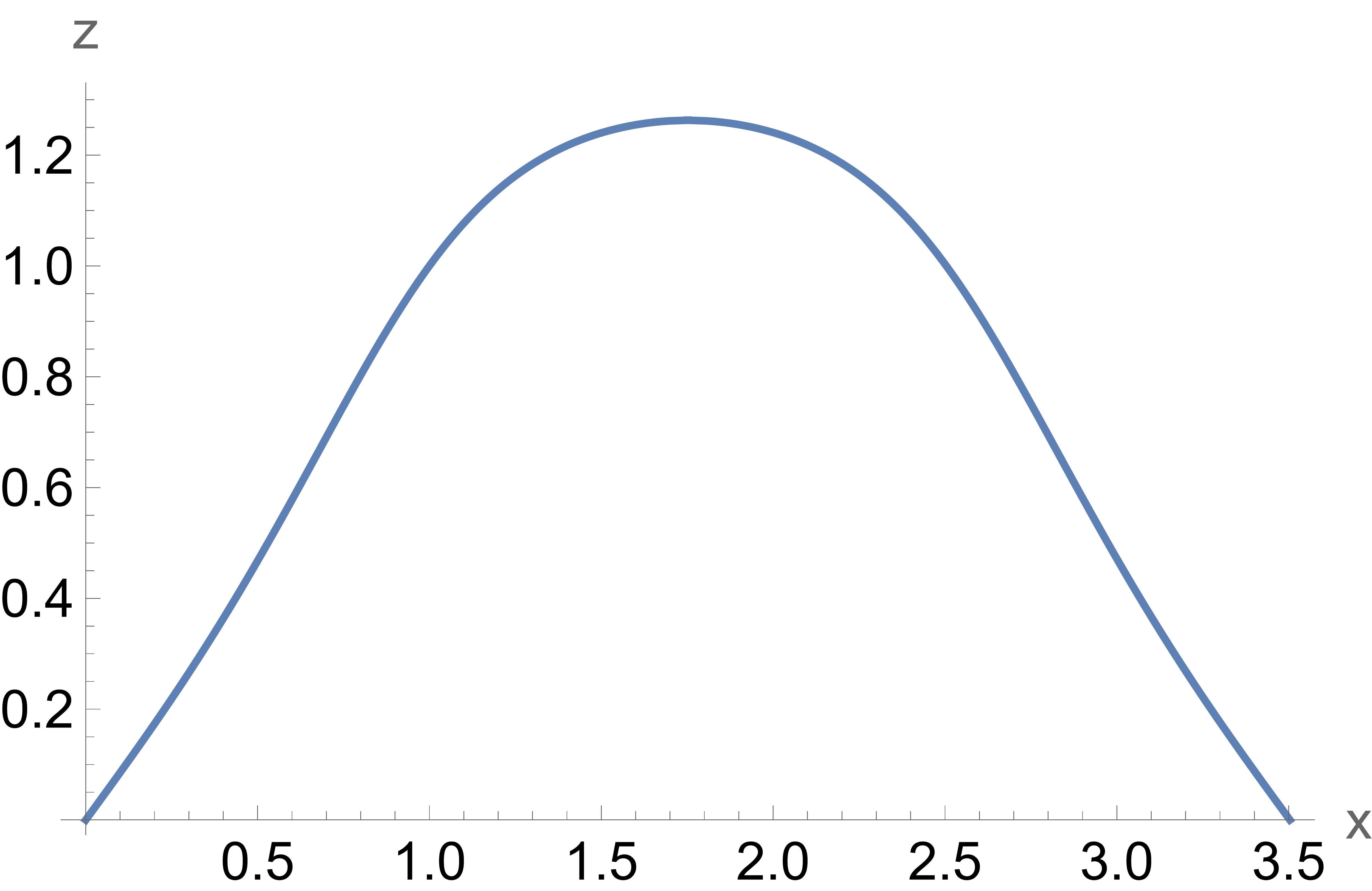}
		\caption{$c_1=1$}
	\end{subfigure}
	\caption{Geometry of EOW branes with $\rho=-1$, $R=1$ and $d=3$ for (a) disconnected phase with $c_1=0$ and (b) connected phase with $c_1=1$. The EOW branes are labeled by blue curves. }
	\label{brane shapes for two c1}
\end{figure}

The bulk metric of the disconnected phase is given by:
\begin{align}\label{sect 3: metric disconnected phase}
ds^2=\frac{\frac{dz^2}{h(z)}+h(z)dx^2+R^2 d\Omega_{d-1}^2}{z^2}, \ \ h(z)=1+\frac{z^2}{R^2},
\end{align}
which is similar to the AdS soliton, except that \( c_1 = 0 \). In this case, the function \( h(z) \) has no zeros, meaning that the coordinate \( z \) is unconstrained and can vary from the boundary at \( z = 0 \) to the deep interior as \( z \to \infty \). The embedding function of the EOW branes satisfies a similar equation to that of the connected phase:
\begin{equation}\label{sect 3: dSa}
	S_a '(z) =\frac{ \mp T}{h(z) \sqrt{(d-1)^2 h(z) - T^2}}.
\end{equation}
where \( a = \ell,r \) indicates the two boundaries of the parallel defects. We take the negative sign for \( a = \ell \) and the positive sign for \( a = r \). Importantly, since the denominator is non-zero—specifically, \( (d-1)^2 h(z) - T^2 = (d-1)^2 (\text{sech}^2(\rho) + \frac{z^2}{R^2}) > 0 \)—there is no turning point where $S_a '(z)\to \pm \infty$. It is the primary technical reason for the branes' disconnection. See Fig. \ref{brane shapes for two c1} for the geometry of EOW branes in the disconnected and connected phases.
In the remainder of this subsection, we set $R=1$. The dependence on $R$ can be restored by replacing $L$ with $L/R$ and $x$ with $x/R$.

To obtain a smooth geometry free of brane intersections, as illustrated in Fig. \ref{brane shapes for two c1} (a), the width \( L \) must be sufficiently large in the case of negative brane tensions. This condition is expressed as: 
\begin{align}\label{sect 3: smallest L disconnected} \text{disconnected}:\  L \ge 2 \int_0^{\infty} S_\ell'(z) \, dz = -2\rho, \ \ \text{for} \ \rho < 0. 
\end{align}
In contrast, for the connected phase, we have:
\begin{align}\label{sect 3: smallest L connected}
\text{connected}:\ 	L\le L_{\text{max}}\ge -2\rho,
\end{align}
where the last inequality is derived from (\ref{sect2: L zero c1}) and Fig.~\ref{relation Lc1} in the limit $c_1\to 0$. Note that the disconnected phase with \( c_1 = 0 \) cannot be considered a smooth continuation of the connected phase as \( c_1 \) approaches zero. As shown in (\ref{sect2: L zero c1}), the connected phase has \( L = -2\rho \) when \( c_1 \to 0^+ \), whereas the disconnected phase exists for all \( L \ge -2\rho \).

Combining the results presented above, we find that multiple classical saddles may coexist for the same boundary width \( L \). Let us first consider the case where \( \rho \leq 0 \). For \( 0 < L < -2\rho \), only the connected saddle exists. However, for \( -2\rho < L < L_{\max} \), both the connected and disconnected saddles coexist. As illustrated in Fig.~\ref{relation Lc1}, two saddles correspond to a given \( L \) in the connected phase. Thus, in total, there are three saddles present. The physical saddle is identified as the one with the lowest free energy. For \( L > L_{\max} \), only the disconnected saddle remains.

Next, we discuss the case where \( \rho \geq 0 \). In this scenario, the disconnected phase exists for all \( L \geq 0 \). Therefore, for \( 0 \leq L \leq L_{\max} \), three saddles coexist: two from the connected phase and one from the disconnected phase. For \( L > L_{\max} \), only the disconnected saddle remains. It is important to note that \( \kappa_1 = 0 \) in this case, indicating that the Casimir effect vanishes.

To conclude this subsection, we note that by applying the following coordinate transformations:
\begin{align} \label{sect 3: coordinate transformation}
	\hat{z}= \frac{z e^{x}}{\sqrt{1+z^2}}, \qquad
	r= \frac{e^{x}}{\sqrt{1+z^2}},
\end{align}
we can express the metric for the disconnected phase in a more familiar form:
\begin{align}\label{sect 3: gravity dual of a ball}
	\mathrm{d} s^{2}= \frac{\mathrm{d}\hat{z}^{2}+\mathrm{d} r^{2}+r^{2}\mathrm{d}\Omega_{d-1}^{2}}{\hat{z}^{2}}.
\end{align}
It represents the holographic dual of a ball \( B^{d} \) with radius \( R_{1} \) (i.e., \( r \leq R_{1} \)). The corresponding EOW brane is described by
\begin{align} \label{sect 3: ball region}
	r^{2}+(\hat{z}-R_{1}\sinh\rho)^{2}= R_{1}^{2}\cosh^{2}\rho.
\end{align}
For the annular geometry where \( R_{1} \leq r \leq R_{2} \), we can combine two brane components: one that ends at the inner sphere \( r = R_{1} \) and another that ends at the outer sphere \( r = R_{2} \). The relationship between the radii and the width \( L \) is given by:
\begin{align} \label{sect 3: BCFT width}
	L=\log\!\left(\frac{R_{2}}{R_{1}}\right).
\end{align}
The bulk dual of the annular region is given by
\begin{align} \label{sect 3: annular region}
	R_{1}^{2}\cosh^{2}\rho-(\hat{z}-R_{1}\sinh\rho)^{2}\le r^{2}\le R_{2}^{2}\cosh^{2}\rho-(\hat{z}-R_{2}\sinh\rho)^{2}. 
\end{align}

\subsection{Phase transition}

This subsection explores the phase transition between the connected and disconnected phases. When multiple saddle points coexist for a given width \( L \), we select the one with the minimal free energy, which corresponds to the minimal renormalized Euclidean action. For simplicity, we will focus on the case where \( d = 3 \) and \( R = 1 \). In this scenario, the renormalized Euclidean action takes the following form:
\begin{align}  \label{sect 3: Euclidean action}
	I_{\mathrm{ren}} &=
	-\int_{N}\!\mathrm{d}^{4}X\sqrt{G}\,(R_N-2\Lambda)
	-2\int_{M}\!\mathrm{d}^{3}x\sqrt{g}\Bigl(K-2-\frac{1}{2}R_{M}\Bigr) \nonumber\\
	&\quad -2\int_{Q}\!\mathrm{d}^{3}y\sqrt{h}\,(K-T)
	-2\int_{P}\!\mathrm{d}^{2}\hat{x}\sqrt{\sigma}\,(\alpha-\alpha_{0}-K_{M}).
\end{align}
Here $K$ and $K_{M}$ denote extrinsic curvatures, $R_{M}$ is the Ricci scalar on the asymptotic boundary $M$ where the BCFT lives, and $P=\partial M=\partial Q$ is the common edge of $M$ and the brane $Q$. The quantity $\alpha=\arccos(n_{M}\cdot n_{Q})$ is the supplementary angle between the unit normals of $M$ and $Q$, and $\alpha_{0}\equiv\alpha(z=0)$ is its value on the AdS boundary.

Let us first discuss the disconnected phase, where the renormalized Euclidean action can be evaluated analytically, yielding
\begin{align} \label{sect 3: disconnected renormalized action}
	I_{\mathrm{discon}}=16\pi\sinh\rho\,\log\epsilon
	+16\pi\arctan(\sinh\rho)
	-4\pi\sinh\rho\Bigl[1+4\log\!\Big(\frac{2}{\cosh\rho}\Big)\Bigr],
\end{align}
where $\epsilon$ is a UV cutoff. The logarithmic divergent term represents the Weyl anomaly, which is independent of the bulk solutions. Interestingly, the finite part of the disconnected phase does not depend on the width $L$. Let us explain why this is the case. Recall that the disconnected phase corresponds to a pure AdS space, meaning its Euclidean action vanishes after subtracting the contributions from the Weyl anomaly and the EOW branes. Next, we will discuss the renormalized Euclidean action of the connected phase, which takes the form:
\begin{align} \label{sect 3: connected renormalized action}
	I_{\mathrm{con}}=16\pi\sinh\rho\,\log\epsilon+ I_{\mathrm{fin}},
\end{align}
where the first term represents the universal Weyl anomaly, and the second term, $I_{\mathrm{fin}}$, can generally be evaluated numerically.

To compare the free energy of the connected and disconnected phases, we analyze the action difference. We define $\Delta I$ by subtracting the contributions from
disconnected phase in the action:
\begin{align} \label{sect 3: con discon dI 0}
	\Delta I=I-I_{\mathrm{discon}}. 
\end{align}
Then, we have $\Delta I_{\mathrm{discon}}=0$ and
\begin{align} \label{sect 3: con discon dI}
	\Delta I_{\mathrm{con}}=I_{\mathrm{con}}-I_{\mathrm{discon}}. 
\end{align}
For the tensionless case, we derive an analytical expression:
\begin{align} \label{sect 3: Iren con 0}
	\Delta I_{\mathrm{con}}|_{\rho=0}=\frac{4\pi L}{z_{\mathrm{max}}}\Bigl(1-\frac{1}{z_{\mathrm{max}}^{2}}\Bigr),
\end{align}
where
\begin{align} \label{sect 3: connected 0 L}
L=\frac{2 \pi  z_{\max }}{3+z_{\max }^2}.
\end{align}
Here, we have used \( L = \beta/2 \) and \( z_{\max} = z_h \) for \( \rho = 0 \), with \( \beta \) being the bulk period defined in Eq.~\eqref{sect 2: beta}. From $\Delta I_{\mathrm{con}}|_{\rho=0}=0$, we derive $z_{\mathrm{max}}=1$ and the critical width
\begin{align}\label{sect 3:  critical width}
	L_{\mathrm{crit}}=\frac{\pi }{2},
\end{align}
which is smaller than the maximum width (\ref{sect 2: L0 maximum}) in the connected phase, i.e., $L_{\mathrm{max}}=\pi /\sqrt{3}$. 

From (\ref{sect 3: Iren con 0}) and (\ref{sect 3: connected 0 L}), we can express the action difference in terms of the width:
\begin{align} \label{sect 3: Iren con 1}
	\Delta I_{\mathrm{con}}|_{\rho=0}=\frac{8 \pi  L^2 \left( \pi ^2-2 L^2\pm \pi  \sqrt{\pi ^2-3 L^2}\right)}{\left(\pi \pm\sqrt{\pi ^2-3 L^2}\right)^3},
\end{align}
where the \(\pm\) symbol indicates that there are two saddle points in the connected phase. Refer to Fig. \ref{relation Lc1}, where one value of \(L\) corresponds to two values of \(c_1\), or equivalently, two values of \(z_{\text{max}}\). We differentiate between the two saddle points based on the values of \(c_1\): \(c_1 > c_{1,\mathrm{t}}\) and \(c_1 < c_{1,\mathrm{t}}\). Here, we define \(L(c_{1,\mathrm{t}}) = L_{\text{max}}\). It turns out that \(c_1 > c_{1,\mathrm{t}}\) is the branch with a smaller free energy.

\begin{figure}[htbp]
	\centering
	\includegraphics[width=0.75\linewidth]{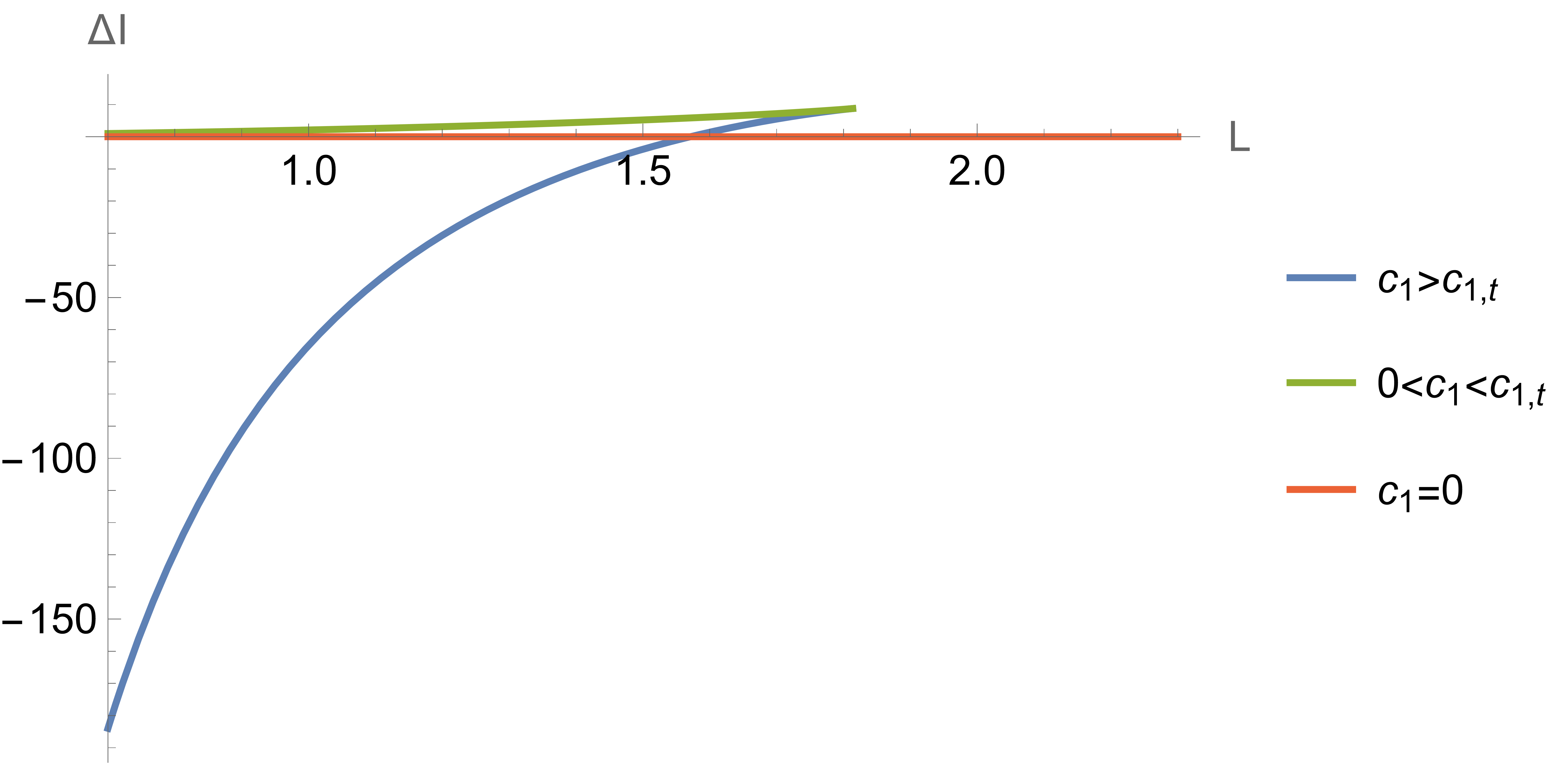}
	\caption{Action difference (\ref{sect 3: Iren con 1}) versus width $L$ for $\rho=0$. The connected phase, characterized by \(c_1 > c_{1,\mathrm{t}}\), dominates when \(L < L_{\mathrm{crit}} = \pi/2\). In contrast, the disconnected phase prevails when \(L > L_{\mathrm{crit}} = \pi/2\).}
	\label{phase transition for rho=0}
\end{figure}

For $\rho=0$, the action difference as a function of \(L\) is depicted in Fig.~\ref{phase transition for rho=0}. It indicates that the connected phase with \(c_1 > c_{1,\mathrm{t}}\) is dominant for:
\begin{align} \label{sect 3: zero T phase 1}
\text{connected phase}:\ 	L\le L_{\mathrm{crit}}=\frac{\pi }{2}.
\end{align}
Conversely, the disconnected phase is dominant for:
\begin{align} \label{sect 3: zero T phase 2}
\text{disconnected phase}:\ 	L\ge L_{\mathrm{crit}}=\frac{\pi }{2}.
\end{align}
We have also evaluated the action difference of the connected phase with $0<c_{1}<c_{1,\mathrm{t}}$; it turns out to be the highest among the three. Thus this phase does not determine the physical phase.

\begin{figure}[htbp]
	\centering
	\includegraphics[width=0.75\linewidth]{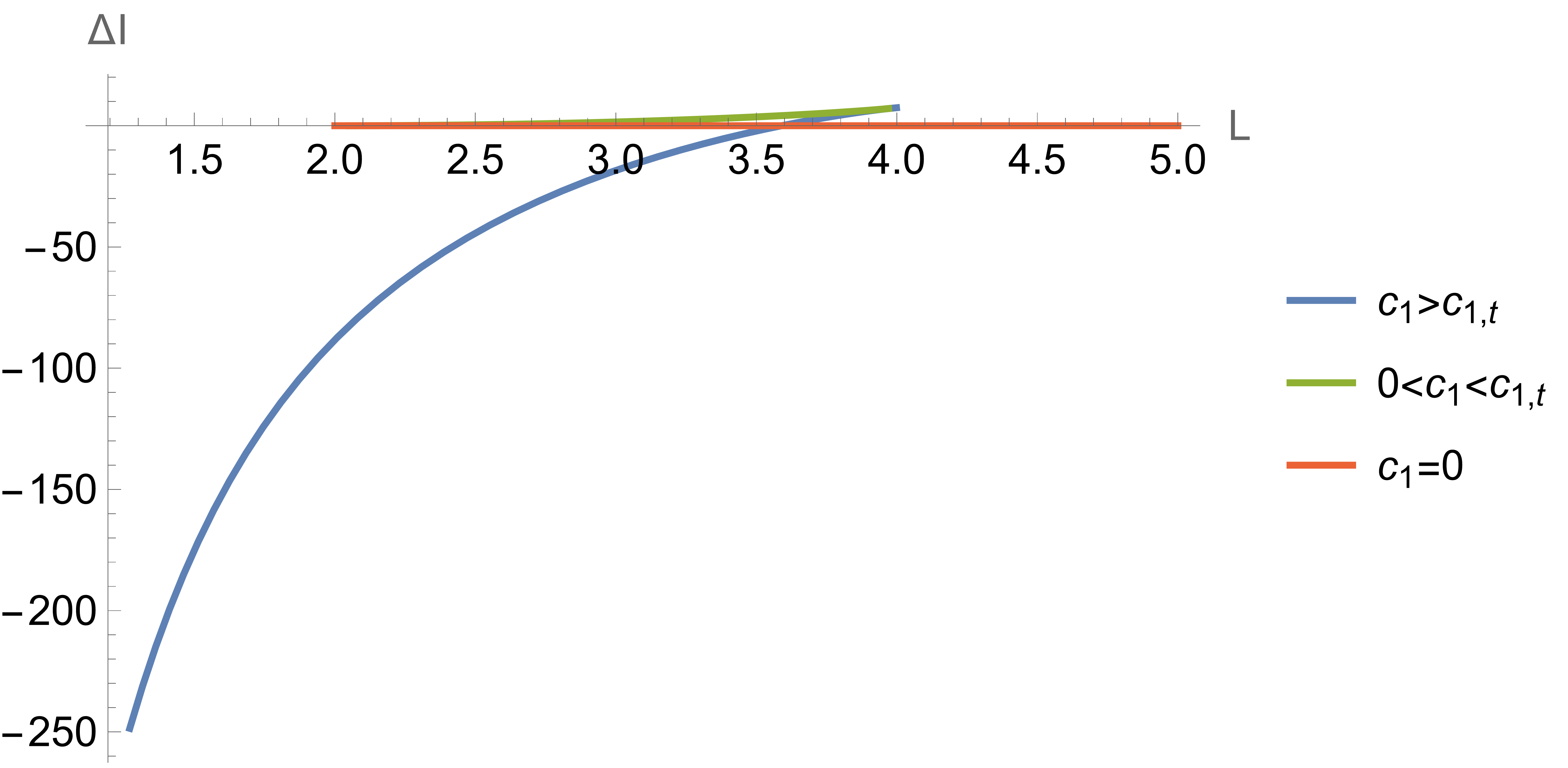}
	\caption{Action difference (\ref{sect 3: con discon dI 0}) versus width $L$ for $R=1$ and $\rho=-1$. The connected phase (blue curve) with \(c_1 >c_{1,\mathrm{t}}\) dominates for $L<L_{\mathrm{crit}}\approx 3.59$, while the disconnected phase (red curve) takes over for $L>L_{\mathrm{crit}}\approx 3.59$. Additionally, the connected phase with \(0 < c_1 < c_{1,\mathrm{t}}\) (green curve) has the highest free energy and is, therefore, thermodynamically unfavorable.}
	\label{phase transition for rho=-1}
\end{figure}

The case for \(\rho \neq 0\) is similar. For example, refer to Fig.~\ref{phase transition for rho=-1} with \(\rho = -1\). In this case, we have \(L_{\mathrm{crit}} \approx 3.59 < L_{\mathrm{max}} \approx 3.97\). For small widths where \(L < L_{\mathrm{crit}}\), the connected phase with \(c_{1} > c_{1,\mathrm{t}}\) dominates. When \(L\) exceeds the critical value \(L_{\mathrm{crit}}\), the disconnected phase becomes dominant. Similarly to the tensionless case, the connected phase with \(c_{1} < c_{1,\mathrm{t}}\) yields the highest free energy, which makes it the physically disfavored phase.

 To conclude this section, we plot $L_{\mathrm{crit}}$ and $L_{\mathrm{max}}$ as functions of $\rho$. It illustrates that $L_{\mathrm{crit}}$ is smaller than $L_{\mathrm{max}},$ but closer to it. 
Additionally, both $L_{\mathrm{crit}}$ and $L_{\mathrm{max}}$ decrease with the tension parameter $\rho$.

\begin{figure}[htbp]
	\centering
	\includegraphics[width=0.75\linewidth]{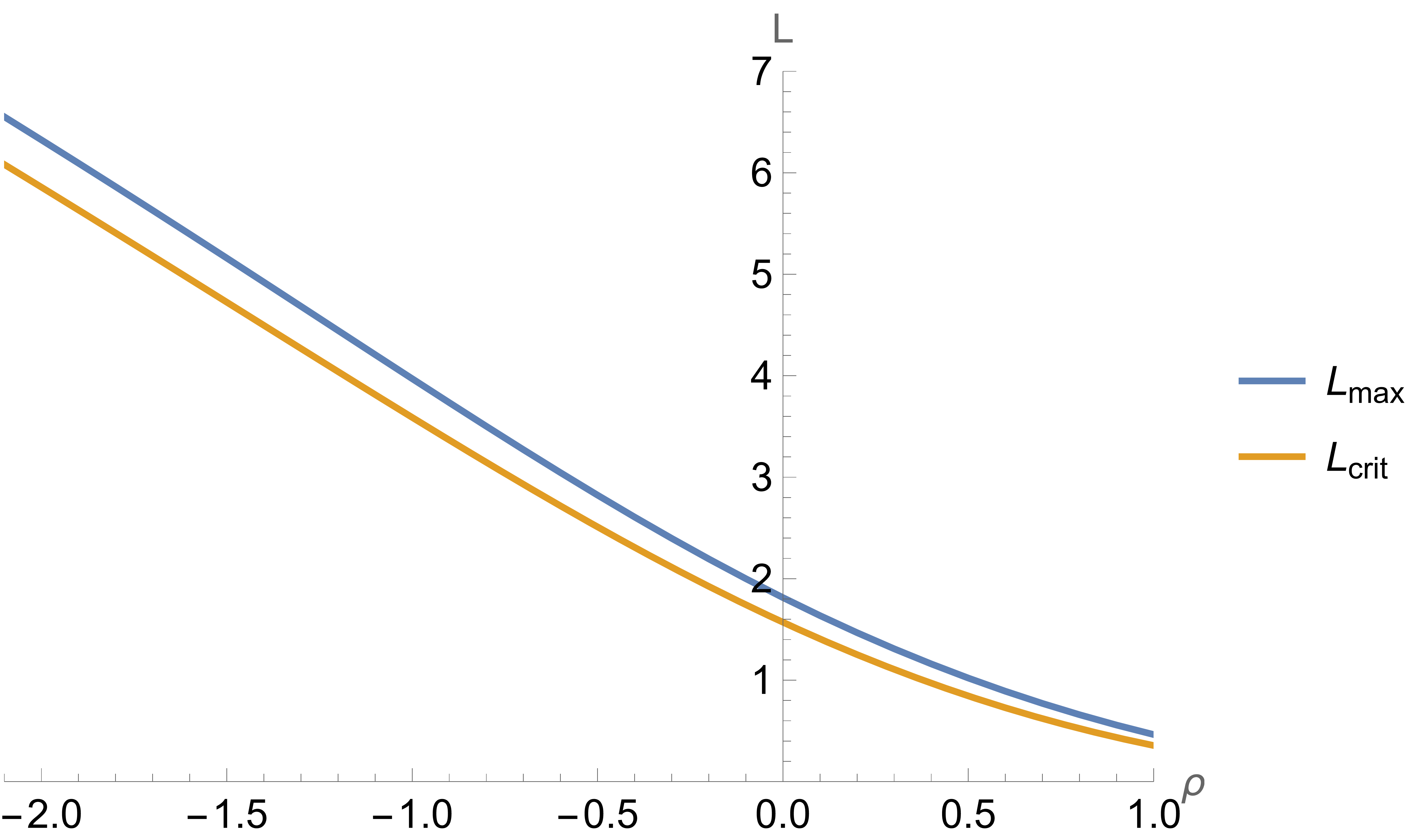}
	\caption{
		The widths $L_{\mathrm{max}}$ and $L_{\mathrm{crit}}$ as functions of the tension $\rho$. Both quantities decrease with increasing $\rho$, and the inequality $L_{\max} \gtrsim L_{\mathrm{crit}}$ is maintained. 
	}
	\label{Lmax and Lcrit}
\end{figure}

In summary, this section examines the phase transition of the gravity dual associated with parallel spherical defects. For small widths, the system is dominated by the AdS soliton with connected EOW branes. In contrast, for larger widths, the system is dominated by the AdS space with disconnected EOW branes. Consequently, a first-order phase transition occurs in the free energy, resulting in a discontinuity in the Casimir pressure at the critical point. Notably, the holographic Casimir effect disappears in the disconnected phase, highlighting a significant difference between strongly coupled theories and free-field theories. For discussions on the Casimir effect in free theory, please refer to Section 5. It's important to emphasize that no such phase transition related to the width \( L \) exists for holographic parallel plane and hyperbolic defects \cite{Takayanagi:2011zk, Miao:2024ddp}. Technically, this difference arises from the variations in the metric function, given by \( h(z) = 1 + k \frac{z^2}{R^2} - c_1 z^d \), where \( k = (0, -1, 1) \) corresponds to plane, hyperbolic, and spherical defects, respectively. Physically, we view the holographic phase transition of parallel spherical defects as a key prediction of this paper.

\section{Censorship principle}

In this section, we discover that various censorship principles impose novel constraints on the holographic Casimir effect. 
They prohibit the singular Casimir effect and clarify why the Casimir force is attractive when the same boundary conditions are applied to both parallel surfaces. Additionally, it provides insight into why the AdS soliton is a reasonable gravitational dual for parallel defects, despite its puzzling properties. Specifically, while the soliton exhibits periodicity in the bulk, it lacks periodicity at the AdS boundary. 
For simplicity, we will use holographic parallel spherical defects as an example in the following discussion. The conclusions drawn here also apply to parallel plane and hyperbolic defects.

\subsection{Cosmic censorship}

Let us first explain why the vacuum of a parallel spherical defect is dual to an AdS soliton. Based on an analysis of symmetries, the Weyl anomaly, and energy conservation, we find that the vacuum expectation values of stress tensors take the form (\ref{Intro: Casimir effect}) for parallel defects in the Introduction. With the CFT stress tensors  (\ref{Intro: Casimir effect}),  we can reconstruct the asymptotically AdS spacetime in the Fefferman–Graham gauge. For further details, see \cite{deHaro:2000vlm}. It turns out that the asymptotic metric near the AdS boundary is equivalent to that of the AdS soliton (\ref{sect 2: holo sphere defect}). This equivalence is most easily observed for odd dimensions \(d\), where the bulk Weyl anomaly vanishes. For simplicity, we focus on \(d=3\) below, but the conclusions apply to general dimensions.

The discussions above indicate that the vacuum of parallel spherical defects is dual to the AdS-soliton (\ref{sect 2: holo sphere defect}) with the metric function 
\begin{align} \label{sect 4: h c1}
h(z)=1+z^2-c_1 z^3,
\end{align}
where we focus on \( R = 1 \) and \( d = 3 \). Here, \( c_1 \) is an arbitrary constant related to the Casimir amplitude, denoted as \( \kappa_1 = c_1 L^3 \). In Section 3, we propose that the metric with \(c_1 > 0\) serves as the dominant saddle for small widths \(L \leq L_{\text{crit}}\), while the metric with \(c_1 = 0\) is the dominant saddle for large widths \(L \geq L_{\text{crit}}\). A natural question arises: why do we not consider the case where \(c_1 < 0\)? In fact, the metric with \(c_1 < 0\) results in the smallest free energy among the three cases. Therefore, this option appears to be the most favored gravity dual for parallel spherical defects.

Let us examine the renormalized Euclidean action for the case where \(c_1 < 0\) and \(d = 3\). For simplicity, we will focus on the scenario involving tensionless EOW branes, meaning \(\rho = 0\). The analysis for cases with tension does not alter the conclusion. 
From (\ref{sect 3: Euclidean action}) with \(T =2\tanh(\rho)=0\), we derive
\begin{align} \label{sect 4: renormalized action for rho=0 and c1<0}
	I_{\text{ren}}(c_1<0) =4\pi c_1 L,
\end{align}
which is unbounded from below as \(c_1 \to -\infty\). Therefore, this represents the dominant saddle point across all widths. However, it leads to a singular Casimir effect, specifically 
\(\kappa_1 = c_1 L^3 \to -\infty\), which is physically unacceptable.

Fortunately, such a pathological solution is prohibited by the principle of cosmic censorship \cite{Penrose:1969pc, Penrose123}. This principle asserts that nature does not favor naked singularities; all singularities should be hidden behind a horizon. The solution with \(c_1 < 0\) exhibits a naked singularity as \(z \to \infty\). To illustrate this, we consider the Kretschmann scalar for \(d = 3\):
\begin{align} \label{sect 4: curvature combination}
	R_{\mu \nu \rho \sigma} R^{\mu \nu \rho \sigma} =24+12 c_1^2 z^6, 
\end{align}
which diverges as \(z \to \infty\). Moreover, for \(c_1 < 0\), the metric function
\begin{align} \label{sect 4: metric function for negative c1}
	h(z)=1+z^2-c_1 z^3>0
\end{align}
has no zeros, indicating that the singularity is naked rather than concealed behind the horizon (where \(h(z_h) = 0\)). Consequently, we must exclude this case from the set of physical solutions.

The reflection positivity of field theories requires an attractive Casimir force if we impose the same boundary conditions on both parallel surfaces \cite{Bachas:2006ti, Diatlyk:2024qpr}. As a byproduct, we obtain a holographic interpretation of this phenomenon. Recall that the holographic Casimir pressure is given by \(T_{nn} = -(d-1) c_1\). The requirement of cosmic censorship dictates that \(c_1 \ge 0\), thereby resulting in an attractive force between the two parallel defects. It is important to note that although the solution with \(c_1 = 0\) has no horizon, its Kretschmann scalar (\ref{sect 4: curvature combination}) is finite, yielding no naked singularities. Now, let's clarify what it means to impose the same boundary conditions on the two boundaries in AdS/BCFT. The tension \(T\) can be interpreted as the gravitational dual of the boundary conditions since it influences the boundary central charges in the same way as the boundary condition itself. It means that the EOW branes terminating at both boundaries must have the same tension. In the connected phase, where \(c_1>0\), this condition is automatically satisfied because only one EOW brane is present. In contrast, for the disconnected phase with \(c_1=0\), we need to ensure that the two disconnected EOW branes also possess the same tension.

In summary, cosmic censorship prevents the divergent Casimir effect and offers a holographic explanation for the attractive Casimir force within the context of AdS/BCFT.

\subsection{Topological censorship}

Let us discuss an intriguing aspect of the gravity dual of parallel defects. For smaller widths, the dual geometry is represented by a portion of the AdS soliton, as illustrated in Fig. \ref{GravityDual}. To eliminate the conical singularity in the bulk, the coordinate \( x \) must have a period \( \beta \) (\ref{sect 2: beta}). However, in the context of the gravity dual for parallel defects, \( x \) is defined within the range \( 0 \leq x \leq L \) instead of exhibiting periodicity \( x \simeq x + \beta \) at the AdS boundary. This discrepancy is indeed puzzling.

As illustrated in Fig. \ref{GravityDual}, for negative brane tension \( T \leq 0 \), the gravity dual is represented by the region between the EOW brane (shown as the blue curve) and the AdS boundary (denoted by the green line). In this scenario, the horizon (depicted as the black line) is hidden behind the EOW brane and is not part of the gravity dual. Therefore, the conical singularity at the horizon is irrelevant. Conversely, for positive brane tension \( T \geq 0 \), the gravity dual encompasses the complement of the region with negative brane tension. Here, the horizon is included in the gravity dual, which raises questions about the aforementioned discrepancy.

Interestingly, this issue can be resolved by the topological censorship \cite{Witten:1999xp, Galloway:1999br}, which asserts that disconnected regions of CFTs cannot correspond to a connected region in the bulk if the null energy condition holds. Assuming there is no periodicity in the bulk, and the two dotted vertical lines shown in Fig. \ref{GravityDual} are not identified, the gravity dual with \( T > 0 \) presents a counterexample to the principle of topological censorship: the two disconnected boundary regions (red lines) correspond to a connected bulk region. However, if there is a periodicity in the bulk and the two dotted vertical lines are identified, this creates connections between the two boundary regions (red lines), aligning the geometry with the topological censorship principle. In this scenario, a connected region of CFTs is dual to a connected region in the bulk.

Next, we will discuss the disconnected phase at large widths, where a portion of AdS space with two disconnected EOW branes gives the gravity dual. See Fig. \ref{brane shapes for two c1} (a). Considering the complement of this geometry, the existence of the disconnected EOW branes leads to two disconnected CFT regions that are dual to two disconnected bulk regions, which complies with the principle of topological censorship.

In summary, we propose that topological censorship offers a natural resolution to the discrepancy regarding the AdS soliton: the bulk exhibits periodicity, while the AdS boundary does not.

\section{Story of free theories}

According to \cite{Miao:2024gcq, Miao:2025utb}, the Casimir effect is bounded by the holographic theory with the minimal brane tension. In this section, we will examine this proposal in the context of parallel spherical defects using free theories. A detailed analysis of the Casimir effect for the free scalar field, Maxwell field, and free Dirac field can be found in Appendix A, Appendix B, and Appendix C, respectively. Below, we highlight the key points and compare them with the holographic bound.

The Casimir amplitude \(\kappa_1\) for a conformally coupled free scalar field is given by \eqref{app A: kappa result}, which is valid for both Dirichlet and Robin boundary conditions. The function \(\kappa_1(L)\) is smooth across all distances. As the separation \(L\) approaches zero, it approaches the flat parallel-plate value, whereas it decays to zero as \(L\) tends to infinity. Therefore, unlike in the holographic theory, the free scalar theory does not exhibit a first-order phase transition in the context of the Casimir effect. By using \eqref{app A: kappa result} and the norm of displacement operator \cite{Miao:2018dvm}
\begin{align}
	\label{sect 5: scalar CD}
	C_D=\frac{\Gamma\left(\frac{d}{2}\right)^2}{2\pi^d},
\end{align}
we obtain the ratio of the negative Casimir amplitude to the norm of the displacement operator, given by:
\begin{align}
	\label{sect 5: scalar ratio}
	-\frac{\kappa_1}{C_D}|_{\text{Scalar}}
	= -\frac{\pi^{d/2} L^d}{2\Gamma(d/2)\Gamma(d)}
	\sum_{\ell=0}^{\infty}
	(2\ell+d-2)^2
	\frac{\Gamma(\ell+d-2)}{\Gamma(\ell+1)}
	\frac{1}{e^{L(2\ell+d-2)}-1}.
\end{align}
Recall that $(-\kappa_1/C_D)$ is proportional to the Casimir pressure per degree of freedom.

Since we are interested in conformal field theories, we focus on the Maxwell theory in four dimensions. There are two types of conformally invariant boundary conditions: the electric boundary condition (also known as the perfect-conductor boundary condition) and the magnetic boundary condition. As in the scalar case, both conditions yield the same Casimir effect for parallel spherical defects. We employ both the heat-kernel method and the spectral method to derive 
the normal-normal component of the stress tensor, and both methods yield the same Casimir amplitude \eqref{app B: kappa result}. Together with the norm of displacement operator \cite{Miao:2018dvm}
\begin{align}
	\label{sect 5: Maxwell CD}
	C_D=\frac{6}{\pi^4},
\end{align}
we find
\begin{align}
	\label{sect 5: Maxwell ratio}
	-\frac{\kappa_1}{C_D}|_{\text{Maxwell}}
	=-\frac{\pi^2 L^4}{18}
	\sum_{\ell=1}^{\infty}
	\frac{\ell(\ell+1)(\ell+2)}{e^{2(\ell+1)L}-1}.
\end{align}

\begin{figure}[htbp]
	\centering
	\includegraphics[width=0.48\textwidth]{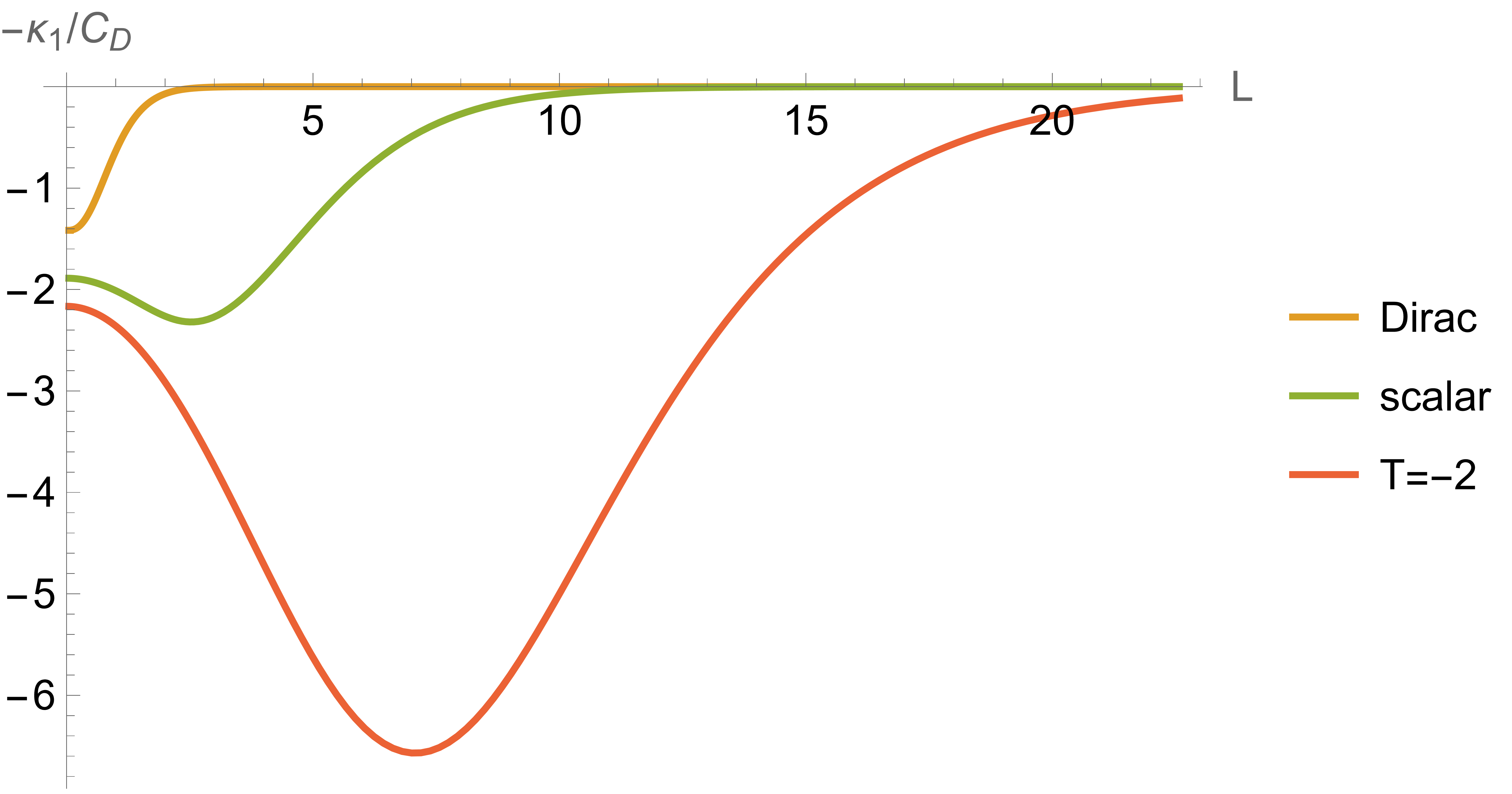}
	\includegraphics[width=0.48\textwidth]{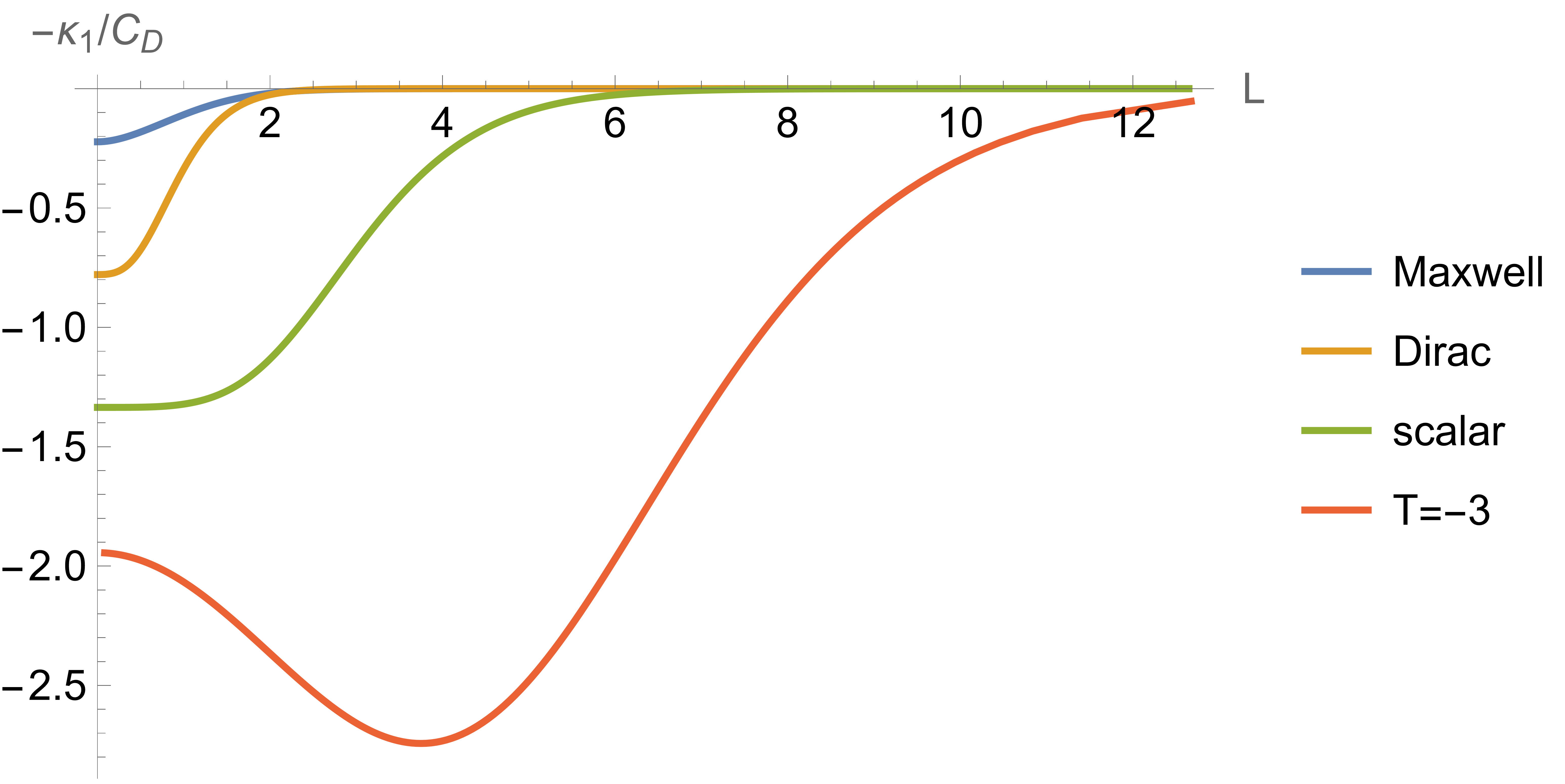}
	\caption{The ratio $-\kappa_1/C_D$ for free theories and the AdS/BCFT with the minimal brane tension $T=-(d-1)$ for $d=3$ (left) and $d=4$ (right). The values of \(-\kappa_1/C_D\) for free fields are positioned above the holographic theory, confirming that free theories obey the holographic bound of the Casimir effect. }
	\label{free fields ratio}
\end{figure}

For the massless free Dirac field, we impose the bag boundary conditions on the two parallel boundaries. By using the spectral method, we derive the Casimir amplitude \eqref{app C: kappa result}. The norm of displacement operator is given by \cite{Miao:2018dvm,McAvity:1993ue}
\begin{align}
	\label{sect 5: Dirac CD}
	C_D=
	\frac{d-1}{4\pi^d}
	\Gamma\left(\frac{d}{2}\right)^2
	2^{\lfloor d/2\rfloor}.
\end{align}
Therefore, we obtain
\begin{align}
	\label{sect 5: Dirac ratio}
	-\frac{\kappa_1}{C_D}|_{\text{Dirac}}
	=-\frac{4\pi^{d/2}L^d}{(d-1)\Gamma(d/2)\Gamma(d)}
	\sum_{\ell=0}^{\infty}
	\frac{\Gamma(\ell+d-1)}{\Gamma(\ell+1)}
	\frac{\ell+\frac{d-1}{2}}
	{e^{2L(\ell+\frac{d-1}{2})}+1}.
\end{align}

The comparison of the ratio $(-\kappa_1/C_D)$ for $d=3$ and $d=4$ is shown in Fig.~\ref{free fields ratio}.  The free field values of $(-\kappa_1/C_D)$ lie above the holographic lower bound with the minimal tension $T=-(d-1)$. It confirms that free theories comply with the holographic bound of the Casimir effect. In the short distance limit, all free field results approach their flat parallel plate values because the local geometry near the two boundaries becomes flat. As the separation increases, the magnitude of the Casimir interaction decreases, and the ratios approach zero. Notably, the ratio (\(-\kappa_1/C_D\) ) is continuous and does not exhibit any phase transition for free theories. This behavior stands in contrast to holographic theories, which exhibit a first-order phase transition for \(T > -(d-1)\). It highlights a significant distinction between free theories and strongly coupled theories.

\section{Conclusions and Discussions}

This paper investigates the holographic Casimir effect for parallel spherical defects. At small widths, the gravity dual is dominated by the AdS soliton with connected EOW branes. In contrast, at larger widths, the dominant phase is the AdS space with disconnected EOW branes. Notably, the Casimir effect exhibits a first-order phase transition 
as the width increases and disappears in the disconnected phase. This behavior differs significantly from that observed in free theories, underscoring a key distinction between free and strongly coupled theories. It is important to highlight that this holographic phase transition is unique to parallel spherical defects and does not occur in parallel plane or hyperbolic defects. It is one of the key findings of this paper. Additionally, we confirm that free theories satisfy the holographic bound of the Casimir effect, further supporting the proposal that the AdS/BCFT with minimal brane tension sets the lower bound of the Casimir effect \cite{Miao:2024gcq, Miao:2025utb}.

Interestingly, we find that cosmic censorship provides a holographic explanation for why the Casimir force is always attractive when the same boundary conditions are applied to both parallel surfaces. The repulsive Casimir force is associated with a bulk spacetime containing a naked singularity, which should be ruled out. Additionally, we discover that topological censorship offers a natural resolution to the discrepancy regarding the AdS soliton: the bulk exhibits periodicity, whereas the AdS boundary does not. These are the main physical findings of this paper. 

So far, studies of the holographic Casimir effect have mainly focused on symmetric configurations, such as parallel plane, hyperbolic and spherical defects. It is intriguing to explore cases with lower symmetry. Surprisingly, the gravity dual of the relatively symmetric case of a ball remains unknown. The metric for the ball is given by:
\begin{align}\label{sect 6: ball}
ds^2=-dt^2+dr^2+r^2 d\Omega^2,\ \ \ r\le R,
\end{align}
where the asymmetry arises from the time component. According to \cite{Miao:2017aba}, the Casimir effect takes a universal form near the boundary
\begin{align}\label{sect 6: Tij}
\langle T_{ij} \rangle \sim C_D \frac{\bar{k}_{ij}}{(R-r)^3}+...
\end{align}
Here, \(\bar{k}_{ij}\) represents the traceless components of the extrinsic curvatures, which are nonzero because of the asymmetry in the time component. In contrast, for parallel defects, we have \(\bar{k}_{ij} = 0\). The non-trivial Casimir effect (\ref{sect 6: Tij}) suggests that the ball cannot be dual to a simple geometry, such as AdS space or an AdS soliton. It is a primary reason why the gravitational dual of a ball has yet to be identified. We intend to address this interesting problem in the future. Another motivation for studying this problem is that there remains debate over whether the Casimir force acting on a spherical shell is attractive or repulsive \cite{Milton:2004ya,Boyer:1968uf,Bowers:1998nr,Kenneth:2006vr}. Holographic studies can provide insights to clarify this controversy.

It is also interesting to study additional aspects of the holographic phase transition, particularly its temperature dependence. The investigation of holographic entanglement entropy and its various generalizations in AdS/BCFT is also a compelling area of research.

\acknowledgments

We thank J. B. Wu for pointing out the discrepancy concerning the AdS soliton, which led us to find the resolution by employing topological censorship. We acknowledge the supports from National Natural Science Foundation of China (NSFC) grant (No.12275366). Miao thanks Yukawa Institute for Theoretical Physics at Kyoto University, where this work was improved during ``YITP-IAS workshop: Interfaces $\&$ Symmetry"  (YITP-I-25-04).


\appendix

\section{Casimir effect of free scalar}\label{app A}

This appendix explores the Casimir effect for parallel spherical defects in a conformally coupled free scalar field, utilizing the heat-kernel method \cite{Vassilevich:2003xt}. The metric of the manifold $M = [0,L] \times S^{d-1}$ is given by
\begin{align}\label{app A: metric}
	ds_M^2 = dx^2 + R^2 d\Omega_{d-1}^2=dx^2 + R^2 h_{ab} d\Omega^a d\Omega^b, \qquad 0\le x\le L,
\end{align}
where $R$ is the radius of the sphere, we set $R=1$ for simplicity, and $L$ is the width of the interval.
The Euclidean action of the conformally coupled free scalar field is given by:
\begin{align}\label{app A: scalar action}
	I = \frac{1}{2} \int_M d^{d} X \sqrt{g_M} \left( \nabla_i \phi \nabla^i \phi + \xi R_M \phi^2 \right)
	+ \int_{\partial M} d^{d-1} \Omega \sqrt{\sigma} \,\xi K_M \phi^2,
\end{align}
where \( \xi = \frac{d-2}{4(d-1)} \) is the conformal coupling constant, \( R_M = (d-1)(d-2) \) is the Ricci scalar, and \( K_M = 0 \) represents the extrinsic curvature. There are two types of conformally invariant boundary conditions:
\begin{align}\label{app A: scalar boundary condition}
	\begin{split}
		&\text{Dirichlet BC} : \phi|_{\partial M}=0,\\
		&\text{Robin BC} : \ \ (\nabla_n + 2\xi K_M)\phi|_{\partial M}=0,
	\end{split}
\end{align}
where $n$ denotes the normal direction. 

The heat kernel satisfies the equation of motion (EOM)
\begin{align}\label{app A: heat kernel EOM}
	\partial_s K(s;X,X') -\Big(\nabla^j \nabla_j -\xi R_M \Big) K(s;X,X')=0,
\end{align}
and the boundary condition (BC)  (\ref{app A: scalar boundary condition}) along with
\begin{align}\label{app A: heat kernel initial condition}
	\lim_{s\rightarrow 0} K(s;X,X') =\frac{\delta^{(d)} (X-X')}{\sqrt{g_M}} .
\end{align}
By solving the above EOM and BCs, we obtain the heat kernel expressed in terms of eigenfunctions:
\begin{align}\label{app A: heat kernel mode sum}
	K(s;X,X')=
	\sum_{\ell=0}^{\infty}\sum_{m=1}^{g_\ell^{(0)}}
	Y_{\ell m}(\Omega)Y_{\ell m}(\Omega')^*
	\begin{cases}
		\displaystyle \sum_{n=1}^{\infty} e^{\lambda s}\frac{2}{L}\sin\!\left(\frac{n\pi x}{L}\right)\sin\!\left(\frac{n\pi x'}{L}\right),\\[0.6em]
		\displaystyle \frac{1}{L}e^{\lambda_0 s}+\sum_{n=1}^{\infty} e^{\lambda s}\frac{2}{L}\cos\!\left(\frac{n\pi x}{L}\right)\cos\!\left(\frac{n\pi x'}{L}\right),
	\end{cases}
\end{align}
where $g_\ell^{(0)}$ is the degeneracy of the harmonics for each $\ell$. The upper entry corresponds to DBC and the lower entry corresponds to RBC. The eigenvalue \( \lambda \) is given by:
\begin{align} \label{app A: eigenvalues}
	\lambda = -\frac{n^2 \pi^2}{L^2} + \lambda_0,
	\qquad
	\lambda_0= -\left(\ell+\frac{d-2}{2}\right)^2 .
\end{align}
It is convenient to rewrite the heat kernel by employing the Poisson summation formula and the properties of spherical harmonics. For Dirichlet BC:
\begin{align} \label{app A: Dirichlet heat kernel}
		K(s;X,X') &= \Bigg[ \sum_{n=1}^{\infty} \frac{2}{L} e^{-\frac{n^2 \pi^2 }{L^2} s}  \sin\!\left( \frac{n \pi x}{L} \right) \sin\!\left( \frac{n \pi x'}{L} \right) \Bigg]  
		\Bigg[ \sum_{\ell=0}^{\infty} \sum_{m=1}^{g_\ell^{(0)}} e^{\lambda_0 s} Y_{\ell m} (\Omega) Y_{\ell m} (\Omega')^* \Bigg] \nonumber \\
		&= \Bigg[ \sum_{n=-\infty}^{\infty} \frac{1}{\sqrt{4\pi s}} \left( e^{-\frac{(x-x'+2nL)^2}{4s}} - e^{-\frac{(x+x'+2nL)^2}{4s}} \right) \Bigg] \nonumber \\
		&\quad \times \Bigg[ \sum_{\ell=0}^{\infty} \frac{\Gamma(d/2)}{2\pi^{d/2}} \frac{2\ell+d-2}{d-2} e^{- \left(\ell+\frac{d-2}{2}\right)^2 s}  C_\ell^{(d/2-1)} (\cos\gamma) \Bigg],
\end{align}
where $C_\ell^{(d/2-1)} (\cos\gamma)$ is the Gegenbauer polynomial, and
\begin{align}\label{app A: geodesic angle}
	\begin{aligned}
		\cos\gamma &= \cos\theta_1 \cos\theta_1' + \sin\theta_1\sin\theta_1'  \\
		&\quad \times \Big[ \cos\theta_2 \cos\theta_2' + \sin\theta_2\sin\theta_2'(\cdots + \sin\theta_{d-2} \sin\theta_{d-2}' \cos(\theta_{d-1} - \theta_{d-1}') \cdots ) \Big].
	\end{aligned}
\end{align}
Similarly, for Robin BC:
\begin{align} \label{app A: Robin heat kernel}
		K(s;X,X') &= \Bigg[\frac{1}{L}+\sum_{n=1}^{\infty} \frac{2}{L} e^{-\frac{n^2 \pi^2 }{L^2} s}  \cos\!\left( \frac{n \pi x}{L} \right) \cos\!\left( \frac{n \pi x'}{L} \right) \Bigg]  
		\Bigg[ \sum_{\ell=0}^{\infty} \sum_{m=1}^{g_\ell^{(0)}} e^{\lambda_0 s} Y_{\ell m} (\Omega) Y_{\ell m} (\Omega')^* \Bigg ]   \nonumber \\
		&= \Bigg[ \sum_{n=-\infty}^{\infty} \frac{1}{\sqrt{4\pi s}} \left( e^{-\frac{(x-x'+2nL)^2}{4s}} + e^{-\frac{(x+x'+2nL)^2}{4s}} \right) \Bigg]  \nonumber\\
		&\quad \times \Bigg[ \sum_{\ell=0}^{\infty} \frac{\Gamma(d/2)}{2\pi^{d/2}} \frac{2\ell+d-2}{d-2} e^{- \left(\ell+\frac{d-2}{2}\right)^2 s}  C_\ell^{(d/2-1)} (\cos\gamma) \Bigg].
\end{align}
The free-space heat kernel is given by
\begin{align} \label{app A: free heat kernel}
	K_0(s;X,X')= \frac{1}{\sqrt{4\pi s}}  \, e^{-\frac{(x-x')^2}{4s}}   \Bigg[ \sum_{\ell=0}^{\infty} \frac{\Gamma(d/2)}{2\pi^{d/2}} \frac{2\ell+d-2}{d-2} e^{\lambda_0 s}  C_\ell^{(d/2-1)} (\cos\gamma) \Bigg].
\end{align}
From the heat kernels (\ref{app A: Dirichlet heat kernel}),(\ref{app A: Robin heat kernel}), and (\ref{app A: free heat kernel}), we can derive the renormalized Green function
\begin{align}\label{app A: Green function}
	\begin{aligned}
		\hat{G} (X,X') &=\int_0^\infty  ds \,\Big(K(s;X,X')-K_0(s;X,X')\Big) \\
		&=k_1 \sum_{\ell=0}^{\infty}  C_\ell^{(d/2-1)} (\cos\gamma) \Bigg[\sum_{n=-\infty}^{\infty} \Bigl( e^{-k_2 |x-x'+2nL|} \mp  e^{-k_2 |x+x'+2nL|} \Bigr) - e^{-k_2 |x-x'|} \Bigg],
	\end{aligned}
\end{align}
where the minus sign corresponds to DBC, and the plus sign corresponds to RBC, and 
\begin{align}  \label{app A: k1 k2}
	k_1=\frac{\Gamma(d/2)}{2(d-2) \pi^{d/2}} ,\qquad k_2=\ell+\frac{d-2}{2}.
\end{align}

\begin{figure}[htbp]
	\centering
	\includegraphics[width=0.47\textwidth]{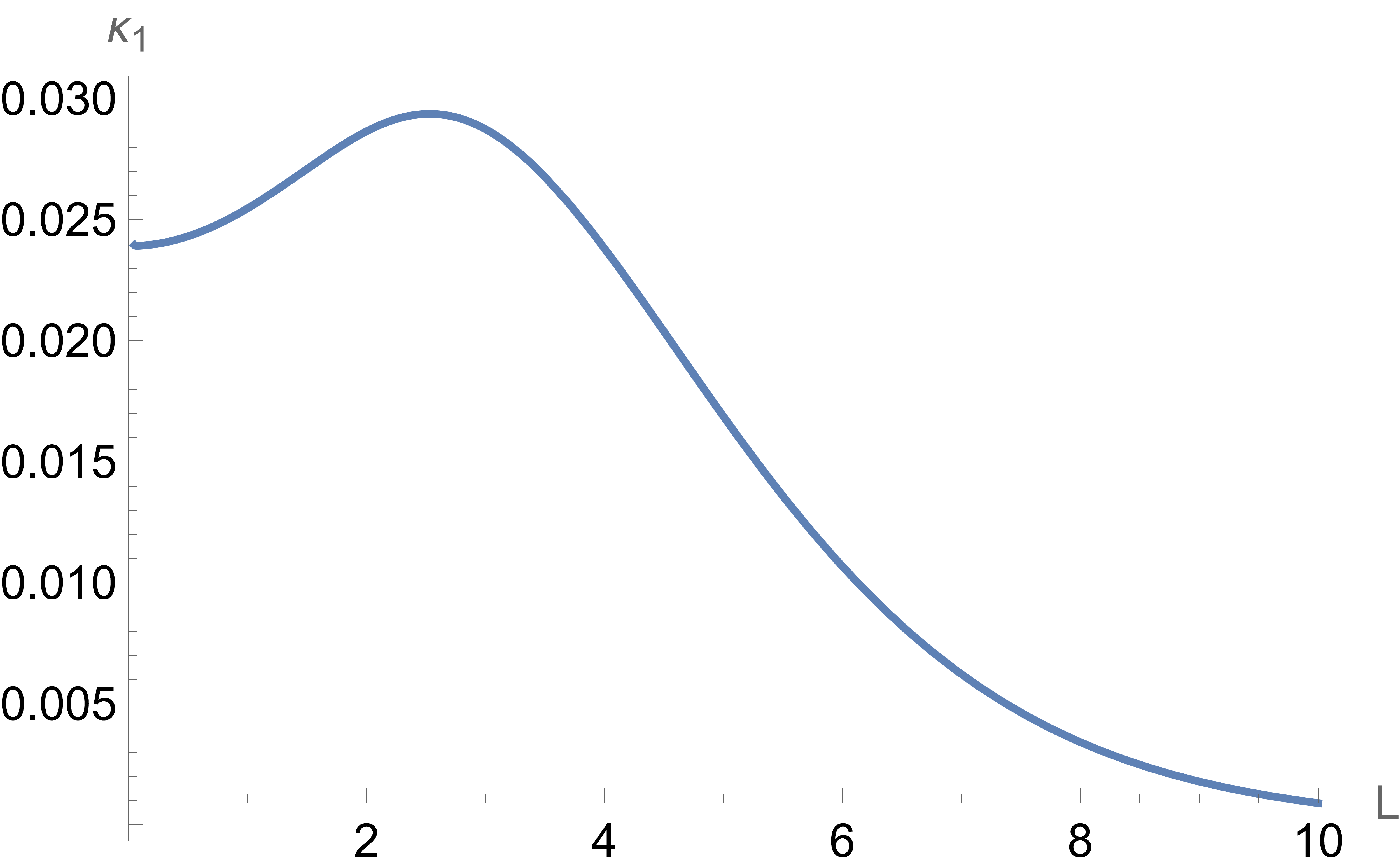}  \includegraphics[width=0.47\textwidth]{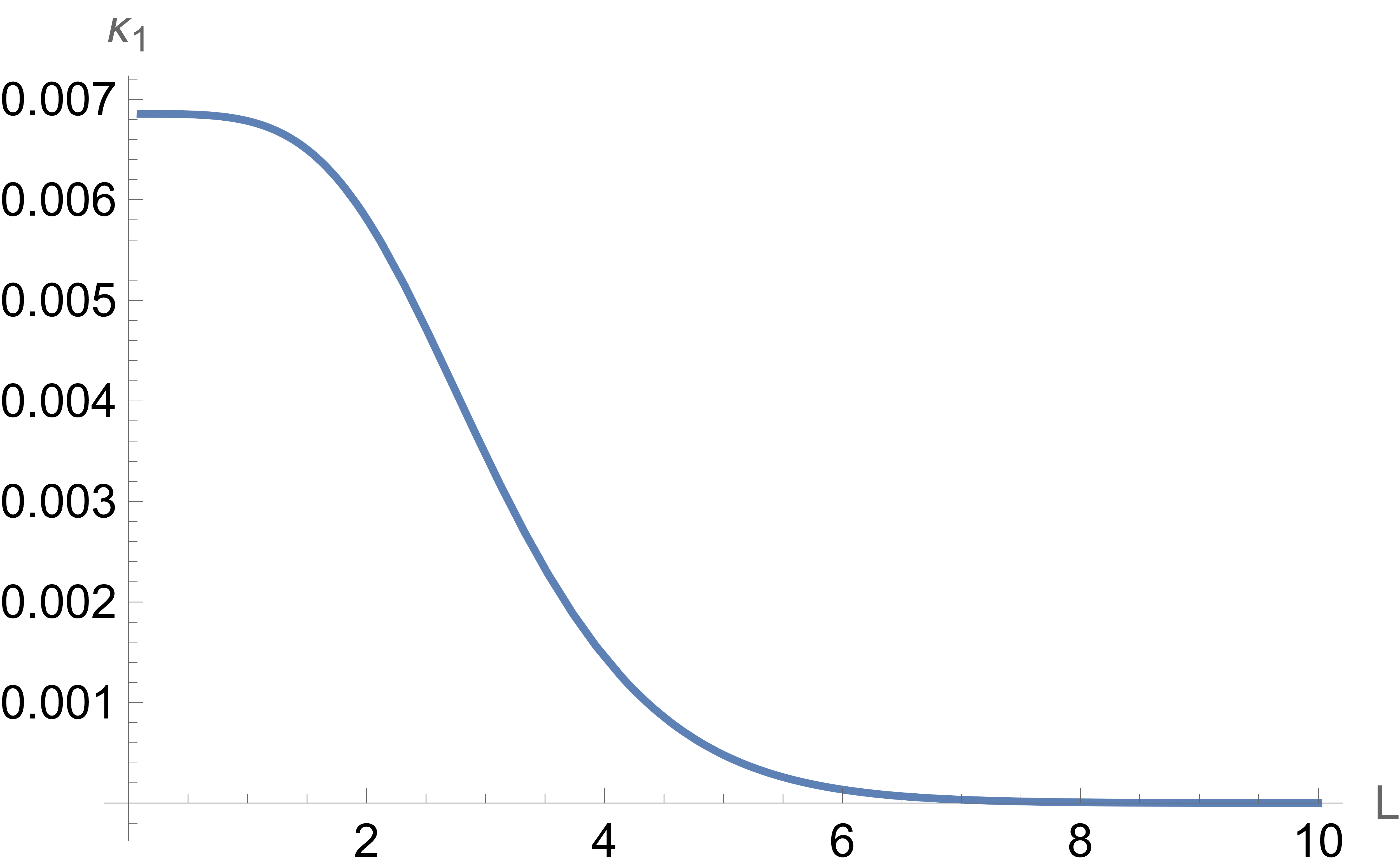} 
	\caption{Casimir amplitude $\kappa_1$ for free scalars with $d=3$ (left) and $d=4$ (right). } 
	\label{app A: kappa figure}
\end{figure}

The renormalized stress tensor is obtained using the following formula:
\begin{align} \label{app A: stress tensor}
	\begin{aligned}
		\langle T_{ij} \rangle   &= \lim\limits_{X\to X'} \Biggl[  (1-2\xi) \nabla_i \nabla_{j'} -2\xi \nabla_i \nabla_j +\Bigl(2\xi-\frac{1}{2} \Bigr)  g_{Mij} \nabla_l \nabla^{l'}  \\
		&\quad + \xi \Bigl( R_{Mij} +\frac{4\xi-1}{2} R_M g_{Mij} \Bigr) \Biggr] \hat{G}(X,X') .
	\end{aligned}
\end{align}
The result is
\begin{align} \label{app A: Txx result}
	\langle T_{xx}  \rangle =- \frac{ \Gamma(d/2)}{4\pi^{d/2}  \Gamma(d-1) }  
	\sum_{\ell=0}^{\infty} (2\ell+d-2)^2 \frac{\Gamma(\ell+d-2)}{ \Gamma(\ell+1)}  
	\frac{1}{e^{L(2\ell+d-2)}-1} .
\end{align}
Comparing the renormalized stress tensor with the expression $\langle T_{xx}  \rangle  =  -(d-1) \kappa_1 /L^d$, we derive the Casimir amplitude:
\begin{align} \label{app A: kappa result}
	\kappa_1 = \frac{ \Gamma(d/2)}{4\pi^{d/2}  \Gamma(d) } L^d 
	\sum_{\ell=0}^{\infty} (2\ell+d-2)^2 \frac{\Gamma(\ell+d-2)}{ \Gamma(\ell+1)}  
	\frac{1}{e^{L(2\ell+d-2)}-1} ,
\end{align}
which is independent of the choice of boundary conditions. Numerical evaluations of this expression for various spacetime dimensions $d = 3$ and $d = 4$ are presented 
in Fig. \ref{app A: kappa figure}.
For a large separation $L$, the Casimir amplitude $\kappa_1$ gradually approaches zero. In contrast, for a small separation $L$, $\kappa_1$ approaches the value corresponding to plane defects, as expected. Interestingly, for $d = 3$, $\kappa_1$ initially increases and then decreases with increasing $L$.

The summation (\ref{app A: kappa result}) can be evaluated analytically in the limit $L \ll 1$. For instance, taking $d=3$, the Casimir amplitude is expressed as
\begin{equation} \label{app A: d3 kappa}
	\kappa_1 = \frac{1}{32\pi} \sum_{\ell=0}^{\infty} (2L) \bigl[(2\ell+1)L\bigr]^2 \frac{1}{e^{(2\ell+1)L}-1}.
\end{equation}
For small $L$, we could approximate the sum by using Euler--Maclaurin formula:
\begin{align} \label{app A: d3 small L}
	\kappa_1 =\frac{\zeta(3)}{16\pi} + \frac{L^2}{192\pi} +\mathcal{O}(L^4) ,
\end{align}
where the leading term $\frac{\zeta(3)}{16\pi}\approx 0.02391$ is consistent with our numerical results. In general dimensions, we find that
\begin{align}\label{app A: small L general d}
	\kappa_1 & \approx \frac{\Gamma(d/2) \,\zeta(d)}{2^d \pi^{d/2}} + \mathcal{O}(L^2), \qquad L \ll 1,
\end{align}
which reproduces the result for parallel plates \cite{Miao:2025utb}. 

\section{Casimir effect of Maxwell theory}\label{app B}

In this appendix, we calculate the Casimir effect of the Maxwell field on the Euclidean product manifold \( M = [0, L] \times S^3 \). We employ two methods: the heat-kernel method and the mode summation method, both of which yield the same Casimir pressure \(T_{nn}\). We restrict attention to four dimensions, since the Maxwell action is conformally invariant only in four dimensions.

Let us start with the Euclidean Maxwell action in the Feynman gauge
\begin{align}
	\label{app B: gauge fixing action}
	I&=\frac14\int_M d^4X\sqrt{g_M}\,F_{ij}F^{ij}
	+\frac12\int_M d^4X\sqrt{g_M}\,(\nabla_i A^i)^2,
\end{align}
where $F_{ij}=\nabla_i A_j-\nabla_j A_i$. After integrating by parts, the quadratic operator $\mathcal D$ can be expressed as:
\begin{align}
	\label{app B: quadratic operator}
	I&=\frac12\int_Md^4X\sqrt{g_M}\,A_i\mathcal D^{ij}A_j,
	\qquad
	\mathcal D^{ij}=-g_M^{ij}\nabla^2+R_M^{ij}.
\end{align}
It leads to the following decompositions:
\begin{align}
	\label{app B: operator decomposition}
	(\mathcal D A)_x&=-\partial_x^2A_x-D^aD_aA_x,\\
	(\mathcal D A)_a&=-\partial_x^2A_a-D^bD_bA_a+2A_a,
\end{align}
where $D_a$ represents the covariant derivative on the unit \( S^3 \).
The boundary term of the action variation is given by:
\begin{align}
	\label{app B: boundary variation}
	\delta I\big|_{\partial M}
	=\int_{\partial M}d^3\Omega\sqrt{\sigma}\,
	\left(n_i F^{ij}+n^j\nabla_k A^k\right)\delta A_j .
\end{align}

Two types of boundary conditions define a well-defined action principle, i.e., $\delta I\big|_{\partial M} = 0$. The first type is the relative, or perfectly conducting, boundary condition, which is given by: 
\begin{align} \label{app B: relative boundary condition} 
A_a\big|_{\partial M} = 0, \qquad \partial_x A_x\big|_{\partial M} = 0. 
\end{align} 
The second type is the absolute boundary condition, represented as: 
\begin{align} \label{app B: absolute boundary condition} 
A_x\big|_{\partial M} = 0, \qquad \partial_x A_a\big|_{\partial M} = 0. \end{align}

Let's briefly review the scalar and vector harmonics on \( S^3 \) \cite{Lindblom:2017maa}, which will be useful for our upcoming calculations. The component \( A_x \) has no \( S^3 \) index, so it behaves as a scalar on \( S^3 \). In contrast, \( A_a \) has one tangential index and must be expanded using vector harmonics.

The normalized scalar harmonics $Y_{\ell m}$ are defined by \cite{Lindblom:2017maa}
\begin{align}
	\label{app B: scalar harmonics}
	-D^aD_aY_{\ell m}(\Omega)
	&=\ell(\ell+2)Y_{\ell m}(\Omega),
	\qquad
	\ell=0,1,2,\ldots,
	\qquad
	m=1,2,\ldots,(\ell+1)^2,
\end{align}
which obeys the following addition theorems:
\begin{align}
	\label{app B: scalar addition theorem}
	\lim\limits_{\Omega\to\Omega'}
	\sum_{m=1}^{(\ell+1)^2}
	Y_{\ell m}(\Omega)Y_{\ell m}^*(\Omega')
	&=\frac{(\ell+1)^2}{2\pi^2},\\
	\lim\limits_{\Omega\to\Omega'}
	\sum_{m=1}^{(\ell+1)^2}
	h^{ab'}D_aY_{\ell m}(\Omega)D_{b'}Y_{\ell m}^*(\Omega')
	&=\frac{\ell(\ell+2)(\ell+1)^2}{2\pi^2} .
\end{align}

A vector field on \( S^3 \) can be decomposed into two parts: a longitudinal part and a transverse, divergence-free part. The normalized longitudinal vector harmonics are defined as follows:
\begin{align}
	\label{app B: longitudinal harmonics}
	Y^L_{a,\ell m}(\Omega)
	&=\frac{D_aY_{\ell m}(\Omega)}{\sqrt{\ell(\ell+2)}},
	\qquad \ell=1,2,\ldots,
	\qquad m=1,2,\ldots,(\ell+1)^2,
\end{align}
which obeys the equation of motion (EOM)
\begin{align}
	\label{app B: longitudinal eigenvalue}
	\left(-D^bD_b+2\right)Y^L_{a,\ell m}
	&=\ell(\ell+2)Y^L_{a,\ell m},
\end{align}
and the following trace identity 
\begin{align}
	\label{app B: longitudinal trace identity}
	\lim\limits_{\Omega\to\Omega'}
	\sum_{m=1}^{(\ell+1)^2}h^{ab'}Y^L_{a,\ell m}(\Omega)Y^{L*}_{b',\ell m}(\Omega')
	&=\frac{(\ell+1)^2}{2\pi^2} .
\end{align}
The transverse harmonics $Y^{T,\sigma}_{a,\ell m}$ are defined by
\begin{align}
	\label{app B: transverse harmonics}
	D^aY^{T,\sigma}_{a,\ell m}=0,
	\quad \ell=1,2,\ldots,&
	\quad
	\sigma=\pm,
	\quad
	m=1,2,\ldots,\ell(\ell+2) ,\\
	\label{app B: transverse harmonics eom}
	\left(-D^bD_b+2\right)Y^{T,\sigma}_{a,\ell m}
	&=(\ell+1)^2Y^{T,\sigma}_{a,\ell m}.
\end{align}
The corresponding trace identity is
\begin{align}
	\label{app B: transverse trace identity}
	\lim\limits_{\Omega\to\Omega'}
	\sum_{\sigma=\pm}\sum_{m=1}^{\ell(\ell+2)}
	h^{ab'}Y^{T,\sigma}_{a,\ell m}(\Omega)Y^{T,\sigma*}_{b',\ell m}(\Omega')
	&=\frac{2\ell(\ell+2)}{2\pi^2} .
\end{align}

\subsection{Heat-kernel method}

The vector heat kernel satisfies the EOM
\begin{align}
	\label{app B: heat kernel EOM}
	\left(\partial_s+\mathcal D\right)_i{}^k K_{kj'}(s;X,X')=0,
\end{align}
along with the boundary conditions (BC)  (\ref{app B: relative boundary condition},\ref{app B: absolute boundary condition}), as well as the initial condition:
\begin{align} \label{app B: heat kernel initial condition}
	\lim\limits_{s\to0}K_{ij'}(s;X,X')
	=g_{Mij'}\frac{\delta^{(4)}(X-X')}{\sqrt{g_M}} .
\end{align}
On the manifold \( M = [0, L] \times S^3 \), the heat kernel can be factored into one-dimensional interval heat kernels and scalar or vector harmonic sums on \( S^3 \).

Let $\hat{K}_N$ and $\hat{K}_D$ denote the renormalized one-dimensional heat kernels on the interval $[0, L]$ with Neumann and Dirichlet BCs. After subtracting the free-space term on $\mathbb{R} \times S^3$, these kernels are given by:
\begin{align}
	\label{app B: Neumann heat kernel}
	\hat{K}_N(s;x,x')
	&=\frac{1}{\sqrt{4\pi s}}
	\left[
	\sum_{\substack{n\in\mathbb Z\\ n\neq0}}
	\exp\left(-\frac{(x-x'+2nL)^2}{4s}\right)
	+
	\sum_{n\in\mathbb Z}
	\exp\left(-\frac{(x+x'+2nL)^2}{4s}\right)
	\right],
	\\
	\label{app B: Dirichlet heat kernel}
	\hat{K}_D(s;x,x')
	&=\frac{1}{\sqrt{4\pi s}}
	\left[
	\sum_{\substack{n\in\mathbb Z\\ n\neq0}}
	\exp\left(-\frac{(x-x'+2nL)^2}{4s}\right)
	-
	\sum_{n\in\mathbb Z}
	\exp\left(-\frac{(x+x'+2nL)^2}{4s}\right)
	\right] .
\end{align}
By utilizing $\hat{K}_N$ and $\hat{K}_D$, we can construct the components of the renormalized heat kernel as follows:
\begin{align}
	\label{app B: heat kernel xx}
	K_{xx'}(s;X,X')
	&=
	\sum_{\ell=0}^{\infty}
	\sum_{m=1}^{(\ell+1)^2}
	e^{-\ell(\ell+2)s}
	Y_{\ell m}(\Omega)Y_{\ell m}^*(\Omega') \begin{cases}
		\hat{K}_N(s;x,x') \\
		\hat{K}_D(s;x,x')
	\end{cases},
	\\
	K_{ab'}(s;X,X')
	&=\begin{cases}
		\hat{K}_D(s;x,x') \\
		\hat{K}_N(s;x,x')
	\end{cases}
	 \times \Bigg[
	\sum_{\ell=1}^{\infty}
	\sum_{m=1}^{(\ell+1)^2}
	e^{-\ell(\ell+2)s}
	Y^L_{a,\ell m}(\Omega)Y^{L*}_{b',\ell m}(\Omega')
	\nonumber \\ \label{app B: heat kernel ij} 
	&\quad +
	\sum_{\ell=1}^{\infty}
	\sum_{\sigma=\pm}
	\sum_{m=1}^{\ell(\ell+2)}
	e^{-(\ell+1)^2s}
	Y^{T,\sigma}_{a,\ell m}(\Omega)Y^{T,\sigma*}_{b',\ell m}(\Omega')
	\Bigg],
	\\
	\label{app B: heat kernel mixed}
	K_{xa'}(s;X,X')&=K_{ax'}(s;X,X')=0 .
\end{align}
In each brace, the upper entry corresponds to relative boundary conditions,
whereas the lower entry corresponds to absolute boundary conditions. From the heat kernel presented above, we can derive the renormalized Green function as follows: 
\begin{align} \label{app B: Green function from heat kernel} \hat{G}_{ij'}(X,X') &= \int_0^\infty ds\,K_{ij'}(s;X,X'). 
\end{align}

The explicit point-splitting calculation below shows that the contribution from $K_{xx'}$ and the longitudinal part of $K_{ab'}$ cancel in the final result. Only the two transverse polarizations contribute to the Casimir pressure 
\begin{align}\label{app B: Txx classical}
	T_{xx}
	&=\frac12h^{ab}F_{xa}F_{xb}
	-\frac14h^{ac}h^{bd}F_{ab}F_{cd} .
\end{align}
To evaluate the expectation value, we separate the two gauge fields into \( X \) and \( X' \), replace \( \langle A_i(X) A_{j'}(X') \rangle \) with the renormalized Green function, apply the necessary derivatives, and finally take the coincidence limit \cite{Christensen:1976vb}. In our notation, we arrive at:
\begin{align}
	\label{app B: point splitting formula}
	\begin{aligned}
		\langle T_{xx}\rangle
		&=\lim\limits_{X\to X'}
		\Bigg\{
		\frac12h^{ab'}\left(\partial_x\partial_{x'}\hat{G}_{ab'}
		+D_aD_{b'}\hat{G}_{xx'}
		-\partial_xD_{b'}\hat{G}_{ax'}
		-D_a\partial_{x'}\hat{G}_{xb'}\right)\\
		&\hspace{1.0cm}
		-\frac14h^{ac'}h^{bd'}\left(D_aD_{c'}\hat{G}_{bd'}
		-D_aD_{d'}\hat{G}_{bc'}
		-D_bD_{c'}\hat{G}_{ad'}
		+D_bD_{d'}\hat{G}_{ac'}\right)
		\Bigg\} .
	\end{aligned}
\end{align}
Since $\hat{G}_{xa'}=\hat{G}_{ax'}=0$, the two mixed terms vanish. It is convenient to denote the remaining six terms by
\begin{align}
	\label{app B: six point splitting terms}
	\begin{aligned}
		I_1&=\lim\limits_{X\to X'}\frac12h^{ab'}\partial_x\partial_{x'}\hat{G}_{ab'},
		&
		I_2&=\lim\limits_{X\to X'}\frac12h^{ab'}D_aD_{b'}\hat{G}_{xx'},\\
		I_3&=-\lim\limits_{X\to X'}\frac14h^{ac'}h^{bd'}D_aD_{c'}\hat{G}_{bd'},
		&
		I_4&=-\lim\limits_{X\to X'}\frac14h^{ac'}h^{bd'}D_bD_{d'}\hat{G}_{ac'},\\
		I_5&=\lim\limits_{X\to X'}\frac14h^{ac'}h^{bd'}D_aD_{d'}\hat{G}_{bc'},
		&
		I_6&=\lim\limits_{X\to X'}\frac14h^{ac'}h^{bd'}D_bD_{c'}\hat{G}_{ad'} .
	\end{aligned}
\end{align}

After inserting the Green's function and applying the coincident identities of the harmonics, the contribution from \( K_{xx'} \) cancels the longitudinal contribution from \( K_{ab'} \). The remaining transverse contributions from \( I_1 + I_2 \) yield:
\begin{align}
	\label{app B: I1 I2 result}
	\begin{aligned}
		I_1+I_2
		&=-\frac{1}{4\pi^2}
		\sum_{\ell=1}^{\infty}\ell(\ell+1)(\ell+2)
		\left[
		\sum_{\substack{n\in\mathbb Z\\ n\neq0}}
		e^{-2(\ell+1)L|n|}
		\pm
		\sum_{n\in\mathbb Z}
		e^{-2(\ell+1)|x+nL|}
		\right] .
	\end{aligned}
\end{align}
Here and below, the upper sign corresponds to relative boundary conditions,
whereas the lower sign corresponds to absolute boundary conditions. Similarly, the transverse parts of \( I_3 + I_4 + I_5 + I_6 \) produce:
\begin{align}
	\label{app B: I3 I6 result}
	\begin{aligned}
		I_3+I_4+I_5+I_6
		&=-\frac{1}{4\pi^2}
		\sum_{\ell=1}^{\infty}\ell(\ell+1)(\ell+2)
		\sum_{\substack{n\in\mathbb Z\\ n\neq0}}
		e^{-2(\ell+1)L|n|}\\
		&\quad
		\pm \frac{1}{4\pi^2}
		\sum_{\ell=1}^{\infty}\ell(\ell+1)(\ell+2)
		\sum_{n\in\mathbb Z}
		e^{-2(\ell+1)|x+nL|} .
	\end{aligned}
\end{align}
The \( x \)-dependent terms cancel out between Eqs. \eqref{app B: I1 I2 result} and \eqref{app B: I3 I6 result}. Therefore, the normal pressure is independent of \( x \) in the interior. Using the identity:
\begin{align}
	\label{app B: image sum}
	\sum_{\substack{n\in\mathbb Z\\ n\neq0}}e^{-2(\ell+1)L|n|}
	&=\frac{2}{e^{2(\ell+1)L}-1},
\end{align}
we obtain
\begin{align}
	\label{app B: Txx final heat kernel}
	\langle T_{xx}\rangle
	&=-\frac{1}{\pi^2}
	\sum_{\ell=1}^{\infty}
	\frac{\ell(\ell+1)(\ell+2)}{e^{2(\ell+1)L}-1} ,
\end{align}
which is independent of the choice of boundary conditions. 
For $L\ll1$, the Euler--Maclaurin expansion gives
\begin{align}
	\label{app B: Txx small L}
	\langle T_{xx}\rangle
	&=-\frac{\pi^2}{240L^4}
	+\frac{1}{24L^2}
	-\frac{1}{4\pi^2L}
	+\frac{11}{240\pi^2}
	+\mathcal O(L) .
\end{align}
The leading term agrees with the standard result for flat plates, as expected in the local flat limit where \(L \ll R\).

\subsection{Mode summation method}

We now present an independent derivation based on the physical-mode summation. On \( S^3 \), the physical polarizations are represented by the transverse and divergence-free vector harmonics \( Y^{T,\sigma}_{a,\ell m} \). The normal scalar component and the longitudinal vector component do not represent independent propagating polarizations, consistent with the previously identified cancellation.

For a physical transverse mode, the gauge potential can be expanded as
\begin{align}
	\label{app B: transverse mode expansion}
	A_a(x,\Omega)
	&=f_n(x)Y^{T,\sigma}_{a,\ell m}(\Omega),
	\qquad
	A_x=0 .
\end{align}
The BCs (\ref{app B: relative boundary condition},\ref{app B: absolute boundary condition}) yield
\begin{align}\label{app B: relative modes}
&\text{relative BC}: \ \ f_n(x)=\sqrt{\frac{2}{L}}\sin\left(\frac{n\pi x}{L}\right), \qquad n=1,2,\ldots ,\\
&\text{absolute BC}: \ f_n(x)=
	\sqrt{\frac{2}{L}}\cos\left(\frac{n \pi x}{L}\right), \qquad n=1,2,\ldots .
\label{app B: absolute modes}
\end{align}

Acting in this mode with the vector operator and utilizing \eqref{app B: transverse harmonics eom}, we obtain the following eigenvalue equation
\begin{align}
	\label{app B: transverse eigenvalue equation}
	\mathcal D_a{}^b\left[f_n(x)Y^{T,\sigma}_{b,\ell m}(\Omega)\right]
	&=\lambda_{n\ell}f_n(x)Y^{T,\sigma}_{a,\ell m}(\Omega) .
\end{align}
The corresponding eigenvalue is given by
\begin{align}
	\label{app B: transverse eigenvalues}
	\lambda_{n\ell}
	&=\left(\frac{n\pi}{L}\right)^2+(\ell+1)^2,
	\qquad
	n=1,2,\ldots,
	\qquad
	\ell=1,2,\ldots .
\end{align}
For absolute BC, there is also a zero mode $f_0(x)=1/\sqrt L$. It gives eigenvalue
$\lambda_{0\ell}=(\ell+1)^2$, which is independent of \(L\). Hence it contributes only an \(L\)-independent term to the effective action and we omit this zero mode in the following \(L\)-dependent Casimir part.
For each value of \(\ell\), the total degeneracy of the physical transverse sector is calculated as follows:
\begin{align}
	\label{app B: physical degeneracy}
	g_\ell
	&=\sum_{\sigma=\pm}\sum_{m=1}^{\ell(\ell+2)}1
	=2\ell(\ell+2) .
\end{align}

For a real bosonic Gaussian mode with eigenvalue \(\lambda\), the functional integral yields \(Z_\lambda \propto \lambda^{-1/2}\). Consequently, the Euclidean effective action can be expressed as
\begin{align}
	\label{app B: determinant formula}
	W=-\log Z=\frac12\sum_\lambda\log\lambda+\text{constant} .
\end{align}
Applying this formula to the physical transverse spectrum gives
\begin{align}
	\label{app B: spectral W before sum}
	\begin{aligned}
		W
		&=\frac12\sum_{\ell=1}^{\infty}g_\ell
		\sum_{n=1}^{\infty}
		\log\left[
		\left(\frac{n\pi}{L}\right)^2+(\ell+1)^2
		\right]\\
		&=\sum_{\ell=1}^{\infty}\ell(\ell+2)
		\sum_{n=1}^{\infty}
		\log\left[
		\left(\frac{n\pi}{L}\right)^2+(\ell+1)^2
		\right] .
	\end{aligned}
\end{align}
To perform the sum over \( n \), we utilize the following formula:
\begin{align}
	\label{app B: sinh product}
	\sinh z
	&=z\prod_{n=1}^{\infty}
	\left(1+\frac{z^2}{n^2\pi^2}\right),
	\qquad
	\sum_{n=1}^{\infty}
	\log\left[
	\left(\frac{n\pi}{L}\right)^2
	\right]=\log(2L) .
\end{align}
From this, we obtain:
\begin{align}
	\label{app B: raw effective action}
	\begin{aligned}
		W&=\sum_{\ell=1}^{\infty}\ell(\ell+2)
		\log\left[1-e^{-2(\ell+1)L}\right] \\
		&\quad +\sum_{\ell=1}^{\infty}\ell(\ell+1)(\ell+2)L
		-\sum_{\ell=1}^{\infty} \ell(\ell+2) \log(\ell+1) .
	\end{aligned}
\end{align}
The third term is independent of \( L \). The second term is proportional to \( L \) and represents the contribution from free space \( \mathbb{R} \times S^3 \). In fact, we have:
\begin{align}
	\label{app B: free space subtraction}
	\begin{aligned} 
		W_{\mathbb R\times S^3}
		&=\frac12\sum_{\ell=1}^{\infty}g_\ell L
		\int_{-\infty}^{\infty}\frac{dk}{2\pi}
		\log\left[k^2+(\ell+1)^2\right]\\
		&=\sum_{\ell=1}^{\infty}\ell(\ell+1)(\ell+2)L
		+\text{constant} .
	\end{aligned}
\end{align}
After subtracting this free-space contribution and disregarding \( L \)-independent terms, the interaction part of the Casimir effective action is given by:
\begin{align}
	\label{app B: Casimir effective action}
	W_{\mathrm{Cas}}(L)
	&=\sum_{\ell=1}^{\infty}\ell(\ell+2)
	\log\left[1-e^{-2(\ell+1)L}\right] .
\end{align}

The normal pressure is obtained by differentiating with respect to separation:
\begin{align}
	\label{app B: spectral Txx final}
	\langle T_{xx}\rangle
	&=-\frac{1}{\operatorname{Vol}(S^3)}
	\frac{\partial W_{\mathrm{Cas}}}{\partial L}
	=-\frac{1}{\pi^2}
	\sum_{\ell=1}^{\infty}
	\frac{\ell(\ell+1)(\ell+2)}{e^{2(\ell+1)L}-1} .
\end{align}
This result is consistent with the point-splitting result presented in \eqref{app B: Txx final heat kernel}. By utilizing 
\begin{align}
	\label{app B: kappa definition}
	\langle T_{xx}\rangle
	&=-\frac{3\kappa_1(L)}{L^4},
\end{align}
we finally obtain the Casimir amplitude for the Maxwell field in four dimensions 
\begin{align}
	\label{app B: kappa result}
	\kappa_1(L)
	=\frac{L^4}{3\pi^2}
	\sum_{\ell=1}^{\infty}
	\frac{\ell(\ell+1)(\ell+2)}{e^{2(\ell+1)L}-1} .
\end{align}

\section{Casimir effect of free Dirac field} \label{app C}

In this appendix, we calculate the interaction part of the Casimir effective action for a free massless Dirac field defined on the Euclidean product manifold \( M = [0, L] \times S^{d-1} \). We impose the bag boundary conditions and apply the mode-sum method, as discussed in Appendix B.

Let us quickly establish our notation for the gamma matrices. Denote the gamma matrices in a local orthonormal frame as \(\Gamma^{\hat A}\), where \(\hat x\) represents the direction along the interval and \(\hat a = 1, \ldots, d-1\) represents the tangent directions on \(S^{d-1}\). The curved gamma matrices are given by \cite{Alcubierre:2025wgj}:
\begin{align}
	\label{app C: curved gamma matrices}
	\gamma^x&=\Gamma^{\hat x},
	\qquad
	\gamma^a=e_{\hat b}{}^a\Gamma^{\hat b},
	\qquad
	\{\Gamma^{\hat A},\Gamma^{\hat B}\}=2\delta^{\hat A\hat B}\mathbf 1_{N_d},
	\qquad
	N_d=2^{\lfloor d/2\rfloor} .
\end{align}
For even dimensions \(d = 2m\), let \(\widetilde\Gamma^{\hat a}\) represent a set of gamma matrices for the Clifford algebra tangent to \(S^{d-1}\), acting on \(2^{m-1}\) components. We choose:
\begin{align}
	\label{app C: even gamma matrices}
	\Gamma^{\hat x}&=\sigma_3\otimes\mathbf 1_{2^{m-1}},
	\qquad
	\Gamma^{\hat a}=\sigma_1\otimes\widetilde\Gamma^{\hat a},
	\qquad
	\{\widetilde\Gamma^{\hat a},\widetilde\Gamma^{\hat b}\}=2\delta^{\hat a\hat b}\mathbf 1_{2^{m-1}} .
\end{align}
For odd dimensions \(d = 2m + 1\), the dimension of the sphere \(d-1 = 2m\) is even. In this case, let \(\widetilde\Gamma^{\hat a}\) denote the gamma matrices on \(S^{2m}\) and define the corresponding chirality matrix as:
\begin{align}
	\label{app C: odd chirality matrix}
	\widetilde\Gamma_*&=i^m\widetilde\Gamma^{\hat 1}\widetilde\Gamma^{\hat 2}\cdots\widetilde\Gamma^{\widehat{2m}},
	\qquad
	\widetilde\Gamma_*^2=1,
	\qquad
	\{\widetilde\Gamma_*,\widetilde\Gamma^{\hat a}\}=0 .
\end{align}
We choose:
\begin{align}
	\label{app C: odd gamma matrices}
	\Gamma^{\hat x}&=\widetilde\Gamma_*,
	\qquad
	\Gamma^{\hat a}=\widetilde\Gamma^{\hat a} .
\end{align}
In both cases, $\Gamma^{\hat x}$ is Hermitian, obeys
$(\Gamma^{\hat x})^2=1$. Thus its eigenvalues are $\pm1$. 
Moreover, $\{\Gamma^{\hat x},\Gamma^{\hat a}\}=0$
implies that the two eigenspaces have equal dimensions. We can therefore
diagonalize $\Gamma^{\hat x}$ as $\Gamma^{\hat x}=\sigma_3\otimes \mathbf1_{N_d/2}.$

We use the Hermitian Dirac spectral operator given by
\begin{align}
	\label{app C: Dirac operator}
	D_d&=-i\gamma^i\nabla_i
	=-i\Gamma^{\hat x}\partial_x+D_{S^{d-1}},
	\qquad
	D_{S^{d-1}}=-i\Gamma^{\hat a}e_{\hat a}{}^a\nabla_a^{S^{d-1}} .
\end{align}
The product geometry has no mixed spin connection between the interval and the sphere. Therefore, we have  $\{\Gamma^{\hat x},D_{S^{d-1}}\}=0$ and $[\partial_x,D_{S^{d-1}}]=0$, which results in
\begin{align}
	\label{app C: squared Dirac operator}
	D_d^2&=-\partial_x^2+D_{S^{d-1}}^2 .
\end{align}
The spectral problem therefore factorizes into an interval problem and the intrinsic Dirac spectrum on $S^{d-1}$.

We impose the bag boundary conditions \cite{Chodos:1974je} as follows:
\begin{align}\label{app C: bag BC}
   \Pi_- \psi |_{\partial M}=0, \qquad \bar{\psi} \Pi_- |_{\partial M}=0 ,\qquad \Pi_-=\frac12 (\mathbf1_{N_d}-n_x \sigma_1 \otimes \mathbf1_{N_d/2} )
\end{align}
so that the boundary term $\bar{\psi} \Gamma_n \delta \psi$ in the variation vanishes on the boundary. These boundary conditions then determine the spectrum. The eigenvalue equation for the Dirac field on the unit sphere is given by \cite{Camporesi:1995fb}
\begin{align}
	\label{app C: sphere Dirac spectrum}
	D_{S^{d-1}}\eta_{\ell m}^s&=s\lambda_\ell\eta_{\ell m}^s,
	\qquad
	s=\pm1,
	\qquad
	\lambda_\ell=\ell+\frac{d-1}{2},
	\qquad
	\ell=0,1,2,\ldots .
\end{align}
For each sign branch, the degeneracy is
\begin{align}
	\label{app C: sphere degeneracy}
	g_\ell&=2^{\lfloor(d-1)/2\rfloor}
	\frac{\Gamma(\ell+d-1)}{\Gamma(d-1)\Gamma(\ell+1)} .
\end{align}
Using the spinor eigenmodes on \(S^{d-1}\), we expand the two components of
the \(d\)-dimensional spinor as
\begin{align}
	\label{app C: spinor mode expansion}
	u(x,\Omega)=f(x)\eta^s_{\ell m}(\Omega),\qquad
	v(x,\Omega)=g(x)\eta^s_{\ell m}(\Omega).
\end{align}
The angular dependence is therefore diagonalized by
\(\eta^s_{\ell m}\), and the remaining eigenvalue problem reduces to a
one-dimensional problem for \(f(x)\) and \(g(x)\).
Substituting the BC \eqref{app C: bag BC} into the interval part of the Dirac eigenvalue equation gives the half-integer momenta
\begin{align}
	\label{app C: interval momenta}
	k_n&=\left(n+\frac12\right)\frac{\pi}{L},
	\qquad
	n=0,1,2,\ldots .
\end{align}
Therefore, the eigenvalues of $D_d^2$ are
\begin{align}
	\label{app C: squared eigenvalues}
	\Lambda_{n\ell}^2&=
	\left(n+\frac12\right)^2\frac{\pi^2}{L^2}
	+\left(\ell+\frac{d-1}{2}\right)^2 .
\end{align}
The total multiplicity entering the determinant of $D_d^2$ is
\begin{align}
	\label{app C: total degeneracy}
	G_\ell&=2g_\ell\frac{N_d}{N_{d-1}}
	=2^{\lfloor d/2\rfloor+1}
	\frac{\Gamma(\ell+d-1)}{\Gamma(d-1)\Gamma(\ell+1)} .
\end{align}
Here, the factor $2$ combines the two signs $s=\pm1$ after squaring, while $N_d/N_{d-1}$ accounts for the embedding of intrinsic sphere spinors into the full $d$-dimensional Dirac spinor.

The effective action of Dirac fields can be calculated as follows:
\begin{align}
	\label{app C: determinant expression}
	W&=-\log Z=-\frac12\log\det D_d^2
	=-\frac12\sum_{\ell=0}^{\infty}G_\ell\sum_{n=0}^{\infty}\log\Lambda_{n\ell}^2 .
\end{align}
By using 
\begin{align}
	\label{app C: product identities}
	\prod_{n=0}^{\infty}\left[\left(n+\frac12\right)^2\frac{\pi^2}{L^2}\right]&=2,
	\qquad
	\cosh z=\prod_{n=0}^{\infty}\left(1+\frac{4z^2}{(2n+1)^2\pi^2}\right),
\end{align}
we derive the effective action
\begin{align}
	\label{app C: raw effective action}
	W&=-2^{\lfloor d/2\rfloor}
	\sum_{\ell=0}^{\infty}
	\frac{\Gamma(\ell+d-1)}{\Gamma(d-1)\Gamma(\ell+1)}
	\left[\lambda_\ell L+\log\left(1+e^{-2\lambda_\ell L}\right)\right] .
\end{align}
The term proportional to \(L\) represents the local contribution from the reference geometry \(\mathbb{R} \times S^{d-1}\). After subtracting this term, the interaction part of the Casimir effective action becomes
\begin{align}
	\label{app C: Casimir effective action}
	W_{\mathrm{Cas}}(L)&=
	-2^{\lfloor d/2\rfloor}
	\sum_{\ell=0}^{\infty}
	\frac{\Gamma(\ell+d-1)}{\Gamma(d-1)\Gamma(\ell+1)}
	\log\left[1+e^{-2L(\ell+\frac{d-1}{2})}\right] .
\end{align}
Dividing by \(\operatorname{Vol}(S^{d-1}) = \frac{2\pi^{d/2}}{\Gamma(d/2)}\) gives us the averaged effective action density
\begin{align}
	\label{app C: effective action density}
	\mathcal W_{\mathrm{Cas}}(L)&=
	-\frac{2^{\lfloor d/2\rfloor}\Gamma(d/2)}{2\pi^{d/2}}
	\sum_{\ell=0}^{\infty}
	\frac{\Gamma(\ell+d-1)}{\Gamma(d-1)\Gamma(\ell+1)}
	\log\left[1+e^{-2L(\ell+\frac{d-1}{2})}\right] .
\end{align}
The normal pressure can be calculated as 
\begin{align}
	\label{app C: normal pressure}
	\langle T_{xx}\rangle&=-\frac{\partial\mathcal W_{\mathrm{Cas}}}{\partial L}
	=-\frac{2^{\lfloor d/2\rfloor}\Gamma(d/2)}{\pi^{d/2}}
	\sum_{\ell=0}^{\infty}
	\frac{\Gamma(\ell+d-1)}{\Gamma(d-1)\Gamma(\ell+1)}
	\frac{\ell+\frac{d-1}{2}}{e^{2L(\ell+\frac{d-1}{2})}+1} .
\end{align}
The pressure is negative, indicating an attractive Casimir force. Recall that $\langle T_{xx}\rangle=-(d-1) \kappa_1/L^d$, we finally obtain the Casimir amplitude
\begin{align}
	\label{app C: kappa result}
	\kappa_1(L)&=
	\frac{2^{\lfloor d/2\rfloor}\Gamma(d/2)L^d}{\pi^{d/2}\Gamma(d)}
	\sum_{\ell=0}^{\infty}
	\frac{\Gamma(\ell+d-1)}{\Gamma(\ell+1)}
	\frac{\ell+\frac{d-1}{2}}{e^{2L(\ell+\frac{d-1}{2})}+1} .
\end{align}

To end this appendix, let us examine the behavior of the Casimir amplitude for small values of \( L \). When \( L \ll 1 \), the primary contribution arises from large values of \( \lambda_\ell = \ell + (d - 1)/2 \). Using the leading asymptotic behavior for large \( \lambda_\ell \):
\begin{align}
	\label{app C: leading degeneracy asymptotics}
	\frac{\Gamma(\ell+d-1)}{\Gamma(d-1)\Gamma(\ell+1)}&\sim
	\frac{\lambda_\ell^{d-2}}{\Gamma(d-1)},
\end{align}
and replacing the sum by an integral, we obtain
\begin{align}
		\kappa_1(L) &\sim
	\frac{2^{\lfloor d/2\rfloor-d}(1-2^{1-d})\Gamma(d/2)\zeta(d)}{\pi^{d/2}} .
\end{align}
This is precisely the flat parallel-plate limit for a massless Dirac field with  bag BC \cite{DePaola:1999im}. For $d=3$, the Euler-Maclaurin expansion yields
\begin{align}\label{app C: d3 small L result}
 \kappa_1=\frac{3\zeta(3)}{32\pi}
	-\frac{L^4}{480\pi}+\cdots
\end{align}
For $d=4$, we find
\begin{align}\label{app C: d4 small L result}
 \kappa_1=\frac{7\pi^2}{2880}
	-\frac{L^2}{288}
	-\frac{17 L^4}{2880\pi^2}+\cdots
\end{align}
Note that we have set the radius \( R = 1 \). When considering the radius, we substitute \( L \) with \( L/R \) in the equations above.




\end{document}